\documentclass[aps,prc,twocolumn,superscriptaddress,floatfix,amsmath,amssymb,nofootinbib,longbibliography]{revtex4-2}
\usepackage[utf8]{inputenc}
\usepackage[T1]{fontenc}

\usepackage{siunitx}
\usepackage{xcolor}
\usepackage{graphicx}
\usepackage{scalefnt}
\usepackage{times,txfonts}
\usepackage{nicefrac}
\usepackage{mathtools}
\usepackage{tabularx}
\usepackage{booktabs}
\usepackage{braket}

\usepackage[colorlinks=true,citecolor=blue,linkcolor=blue,urlcolor=blue]{hyperref}
\usepackage{orcidlink}

\newcommand{\be}{\begin{equation}}
\newcommand{\ee}{\end{equation}}

\newcommand{\fmiq}{\, \text{fm}^{-3}}
\newcommand{\mev}{\, \text{MeV}}

\newcommand{\eref}{E_\mathrm{ref}}
\newcommand{\erefBQ}{E_\mathrm{ref}^\mathrm{BQ}}
\newcommand{\erefB}{E_\mathrm{ref}^\mathrm{BQ + 3B}}

\begin{document}

\title{Neutron star crust and outer core equation of state\\ from chiral effective field theory with quantified uncertainties}

\author{H.~G\"ottling\orcidlink{0009-0005-7454-5956}}
\email{hannah.goettling@tu-darmstadt.de}
\affiliation{Technische Universit\"at Darmstadt, Department of Physics, D-64289 Darmstadt, Germany} 
\affiliation{ExtreMe Matter Institute EMMI, GSI Helmholtzzentrum f\"ur Schwerionenforschung GmbH, D-64291 Darmstadt, Germany}

\author{L.~Hoff\orcidlink{0009-0005-4313-5743}}
\email{luis.hoff@stud.tu-darmstadt.de}
\affiliation{Technische Universit\"at Darmstadt, Department of Physics, D-64289 Darmstadt, Germany} 
\affiliation{ExtreMe Matter Institute EMMI, GSI Helmholtzzentrum f\"ur Schwerionenforschung GmbH, D-64291 Darmstadt, Germany}

\author{K.~Hebeler\orcidlink{0000-0003-0640-1801}}
\email{kai.hebeler@tu-darmstadt.de}
\affiliation{Technische Universit\"at Darmstadt, Department of Physics, D-64289 Darmstadt, Germany} 
\affiliation{ExtreMe Matter Institute EMMI, GSI Helmholtzzentrum f\"ur Schwerionenforschung GmbH, D-64291 Darmstadt, Germany}
\affiliation{Max-Planck-Institut f\"ur Kernphysik, Saupfercheckweg 1, D-69117 Heidelberg, Germany}

\author{A.~Schwenk\orcidlink{0000-0001-8027-4076}}
\email{schwenk@physik.tu-darmstadt.de}
\affiliation{Technische Universit\"at Darmstadt, Department of Physics, D-64289 Darmstadt, Germany} 
\affiliation{ExtreMe Matter Institute EMMI, GSI Helmholtzzentrum f\"ur Schwerionenforschung GmbH, D-64291 Darmstadt, Germany}
\affiliation{Max-Planck-Institut f\"ur Kernphysik, Saupfercheckweg 1, D-69117 Heidelberg, Germany}

\begin{abstract} 
We study the order-by-order expansion of the energy per particle of asymmetric nuclear matter up to twice saturation density in chiral effective field theory (EFT) within a Bayesian framework. For this, we develop a two-dimensional Gaussian process (2D GP) that is trained using many-body perturbation theory results based on chiral two- and three-nucleon interactions from leading to next-to-next-to-next-to-leading order (N$^3$LO). This allows for an efficient evaluation of the equation of state (EOS) and thermodynamic derivatives with EFT truncation uncertainties. After benchmarking our 2D GP against Bayesian uncertainties for pure neutron matter and symmetric matter, we study the energy per particle, pressure, and chemical potentials of neutron star matter in $\beta$ equilibrium including EFT uncertainties. We investigate the phase diagram of neutron-rich matter from neutron- to proton-drip and to the uniform phase, including surface and Coulomb corrections. Based on this, we construct EOSs for the inner crust of neutron stars that are consistent with the chiral EFT results for uniform matter at N$^3$LO.
\end{abstract}

\maketitle 

\section{Introduction}

The nuclear equation of state (EOS) is the key quantity for understanding the nature of matter in the interior of neutron stars, for simulations of core-collapse supernovae, and for neutron star mergers (see, e.g., Refs.~\mbox{\cite{Hebeler_2013,
Yasin:2018ckc,Schneider:2019shi,Lattimer2021,Jacobi:2023olu,EOSreview})}. For such applications, information about various thermodynamic quantities, e.g., pressure, energy density, and chemical potentials, is needed over a wide range of densities, proton fractions, and also temperatures. In addition, it is important to provide theoretical uncertainties for these quantities. In this work, we develop a framework that is able to calculate the EOS for arbitrary densities and proton fractions in the nucleonic regime and includes the largest uncertainties from truncations in nuclear forces.

With advances in nuclear theory of the last decades, it has become possible to calculate thermodynamic properties of asymmetric nuclear matter up to about twice nuclear saturation density in a systematic way. These advances have been realized by the derivation of modern nuclear forces from chiral effective field theory (EFT) (see, e.g., Refs.~\mbox{\cite{Epelbaum_et_al_2009,Machleidt_Entem_2011}}) and by powerful many-body developments based on these interactions~\cite{HebelerSchwenk2010,Tews13N3LO,Holt13PPNP,Carb13nm,Hage14ccnm,Coraggio_et_al_2014,PhysRevC.89.064009,PhysRevC.92.015801,Lynn16QMC3N,Dris16asym,Ekst17deltasat,Drischler2019,CarboneSchwenk2019,Lu2020,Keller2021,Keller:2022crb,Marino:2024tfp,Tews:2024owl}. Chiral EFT orders the different interaction contributions according to their importance at low momenta. With this EFT expansion, it is possible to estimate the uncertainties for a given observable due to the neglected contributions from higher orders. While there are other sources of theoretical uncertainties, e.g., from the calibration data used for nuclear forces and from the many-body approach, for the EOS the interaction uncertainties from the EFT truncation are dominant (see, e.g., the recent overview of EOS calculations from different chiral two- and three-nucleon interactions~\cite{Alp:2025wjn}).

Bayesian approaches are an ideal framework for assessing theoretical uncertainties when there is prior information like the EFT expansion. They allow to include new information order-by-order, incorporate prior information, and determine posterior distributions for observables. For this purpose, Gaussian processes (GPs)~\cite{Rasmussen2005} provide a powerful tool as emulators of EOS calculations~\cite{Keller:2022crb}, to extract correlations in the independent variables and to efficiently and robustly compute derivatives of observables, as well as to encorporate EFT uncertainties. The approach of this work is strongly inspired by Refs.~\cite{Melendez:2019izc_EFT,Drischler:2020hwi_EOS,Drischler:2020yad_matter}, in which a GP-Bayesian uncertainty framework was developed for pure neutron matter (PNM) and symmetric nuclear matter (SNM). Here we generalize this work by extending the Gaussian process to two independent variables, i.e., density $n$ and proton fraction $x$, and perform a detailed EFT uncertainty analysis for neutron-rich matter in $\beta$ equilibrium.

\begin{figure*}[t!]
    \centering
    \includegraphics[width=0.8\linewidth]{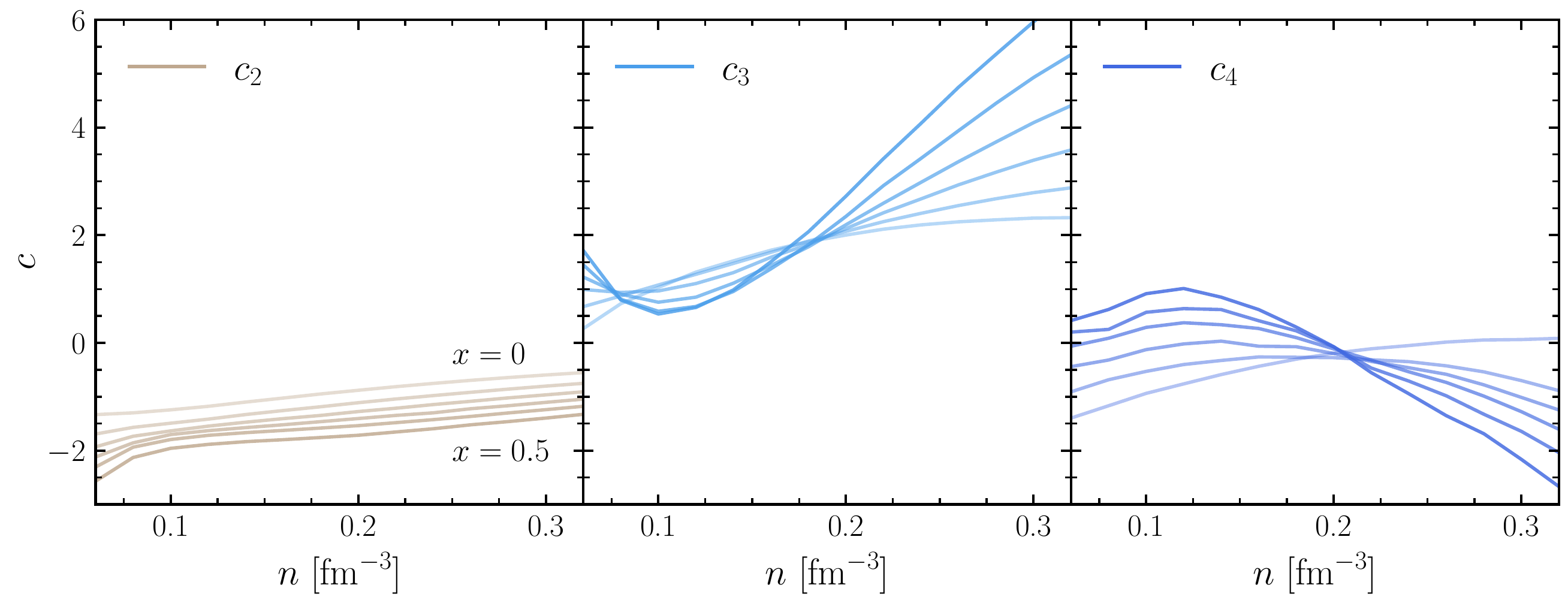}
    \caption{Expansion coefficients at NLO ($c_2$, left), N$^2$LO ($c_3$, middle), and N$^3$LO ($c_4$, right panel) extracted from the asymmetric matter calculations with reference energy $E_\text{ref}^\text{BQ}$ as a function of density $n$. Results are shown for different proton fractions from $x=0$ to $x=0.5$ in steps of $0.1$ (from light to darker).}
    \label{fig:c_n}
\end{figure*}

\begin{figure*}[t!]
    \centering
    \includegraphics[width=0.8\linewidth]{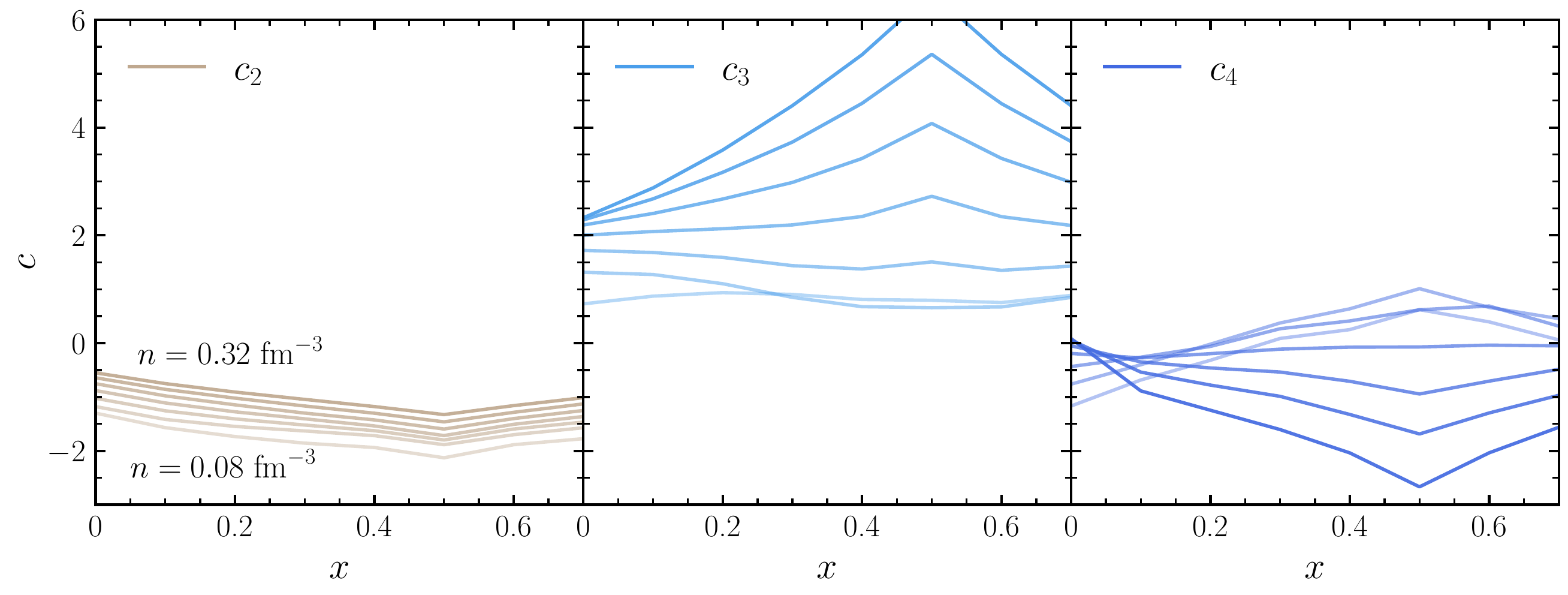}
    \caption{Expansion coefficients analogous to Fig.~\ref{fig:c_n}, here as function of proton fraction $x$ shown for selected fixed densities between $n = 0.08 \fmiq$ and $n = 0.32 \fmiq$ in steps of $0.04 \fmiq$ (from light to darker). Note that the kink at $x=0.5$ is due to asymmetric matter results being on a grid of proton fractions, the actual behavior around $x=0.5$ is smooth.}
    \label{fig:c_x}
\end{figure*}

This paper is organized as follows. In Sec.~\ref{sec:EFT} we study the order-by-order expansion of the energy per particle of asymmetric nuclear matter and discuss the corresponding expansion coefficients based on microscopic calculations from NN and 3N interactions up to N$^3$LO in chiral EFT. In Sec.~\ref{sec:2DGP} we develop a two-dimensional GP for EFT uncertainties as a function of density and proton fraction, discuss training and optimization choices, and provide model-checking diagnostics. In Sec.~\ref{sec:results}, we present the resulting EFT uncertainties for the EOS and related properties, for different proton fractions including matter in $\beta$ equilibrium. This includes the energy per particle, pressure, and chemical potentials. Based on these results, we construct in Sec.~\ref{sec:crust} EOSs for the inner crust of neutron stars including surface and Coulomb corrections, which are consistent with the chiral EFT calculations for uniform matter at N$^3$LO. Finally, we summarize and give an outlook in Sec.~\ref{sec:summary}.

\section{Chiral EFT and order-by-order results}
\label{sec:EFT}

Our results are based on the order-by-order chiral EFT calculations of asymmetric nuclear matter by Keller {\it et al.}~\cite{Keller:2022crb}. The calculations have been performed using many-body perturbation theory up to third order based on the Entem, Machleidt, Nosyk NN interactions~\cite{Entem:2017gor} at different orders with cutoff $450 \mev$ and 3N interactions up to N$^3$LO, fitted to the $^3$H binding energy and the empirical saturation region~\cite{Drischler2019}. In Ref.~\cite{Keller:2022crb}, the EOS and related thermodynamic properties were computed as a function of total number density $n=n_n+n_p$, proton fraction $x = n_p/n$, and temperature $T$. In this work, we focus on the energy per particle at zero temperature and write it explicitly as an EFT expansion of the following form~\cite{Furnstahl:2015rha,Drischler:2020hwi_EOS,Drischler:2020yad_matter}:
\begin{align}
    E(n,x) &= E_{\text{ref}}(n,x) \sum_{\mathnormal{i}=0}^{\infty} c_{\mathnormal{i}}(n,x) \left[ Q (n,x) \right]^{\mathnormal{i}} \, , 
    \label{eq:seriesdefinition} \\
    &= E_k(n,x) + \delta E_k (n,x) \, ,
\end{align}
where $E_{\text{ref}}(n,x)$ denotes a reference energy, $c_{\mathnormal{i}}(n,x)$ are the expansion coefficients, and the expansion parameter $Q(n,x)$ encodes the expected convergence behavior resulting from the EFT power counting. The expansion parameter contains the relevant momentum scale of the considered system and the chiral EFT breakdown scale $\Lambda_b$. In this work, we choose
\begin{equation}
    Q(n,x) = \frac{\bar{k}_{\mathrm{F}} (n,x)}{\Lambda_b} \, ,    
\end{equation}
where the Fermi momentum scale $\bar{k}_{\text{F}} (n,x)$ is taken from
\begin{equation}
    \bar{k}_{\text{F}} (n,x) = \text{max} \bigl( k_{\mathrm{F}}^{n}, k_{\mathrm{F}}^p \bigr) \, ,
\end{equation}
where $k_{\mathrm{F}}^{n}$ and $k_{\mathrm{F}}^p$ are the neutron and proton Fermi momenta, respectively. For the breakdown scale we use \mbox{$\Lambda_b=\SI{600}{MeV}$}~\cite{Drischler:2020hwi_EOS,Drischler:2020yad_matter}.

The contributions up to a given order $k$ are given by $E_k(n,x)$, i.e.,
\begin{align}\label{eq:contributions}
    E_k (n,x) &= E_{\text{ref}}(n,x) \sum_{\mathnormal{i}=0}^k c_{\mathnormal{i}}(n,x) \left[ Q (n,x) \right]^{\mathnormal{i}} \, ,
\end{align}
while $\delta E_k(n,x)$ is the EFT truncation uncertainty at this order, which includes the contributions from all higher orders beyond $k$:
\begin{align}
    \delta E_k (n,x) &= E_{\text{ref}}(n,x)  \sum_{\mathnormal{i}=k+1}^\infty c_{\mathnormal{i}}(n,x) \left[ Q (n,x) \right]^{\mathnormal{i}} \, .
    \label{eq:DeltaEk}
\end{align}
The dimensionful reference energy $E_{\text{ref}}(n,x)$ is introduced to set the scale of the observable, such that the coefficients $c_i$ are dimensionless and of natural size, i.e., of order 1. For the reference energy per particle $E_{\text{ref}}/A$, we follow Ref.~\cite{Drischler:2020hwi_EOS,Drischler:2020yad_matter} (indicated as BQ) and first explore
\begin{equation}
     \frac{\eref(n,x)}{A} = \frac{\erefBQ (n)}{A} = \SI{16}{MeV} \, (n/n_0)^{2/3} \, ,
     \label{eq:ref}
\end{equation}
with nuclear saturation density $n_0=0.16\fmiq$. This relation is inspired by the density dependence of a non-interacting Fermi gas and is adjusted to give a reasonable energy at saturation density. Moreover, this reference scale is overall consistent with leading-order (LO) predictions of PNM and SNM~\cite{Drischler:2020hwi_EOS,Drischler:2020yad_matter}. For PNM the density dependence is also that of the unitary Fermi gas, while the coefficient matches reasonable values of the symmetry energy. Note that this reference energy is independent of the proton fraction~$x$. We will consider a more suitable choice to cover asymmetric matter with arbitrary proton fraction below.

With these choices for the reference energy and the expansion parameter, the expansion coefficients are determined according to Eq.~\eqref{eq:contributions} from the asymmetric matter results of the energy per particle for the available orders. Figure~\ref{fig:c_n} shows the resulting expansion coefficients $c_2(n,x)$, $c_3(n,x)$, and $c_4(n,x)$ at fixed proton fraction as a function of density from $n= 0.06 \fmiq$ to $2 n_0$. We find values of the coefficients in the range $c_i(n,x) \approx [-3,7]$ at all orders. Generally, the values tend to be largest for SNM ($x=0.5$) and increase towards larger densities. The significant change in size from NLO ($c_2$) to N$^2$LO~($c_3$) is due to the contributions of 3N interactions that enter at this order. Their contributions are especially prominent at larger densities. Figure~\ref{fig:c_x} shows the expansion coefficients, but at fixed densities as a function of proton fraction up to $x=0.7$. Due to the approximate isospin symmetry, the results are to a very good approximation symmetric around $x=0.5$.

\section{Two-dimensional Gaussian process for EFT uncertainties}
\label{sec:2DGP}

To estimate the EFT truncation uncertainties, i.e., the value of $\delta E_k(n,x)$, we develop a two-dimensional GP in the vector variable $\mathbf{y} = (n,x)$. This follows the GP-Bayesian uncertainties in the one-dimensional (1D) case, i.e., for PNM or for SNM, from Refs.~\cite{Drischler:2020hwi_EOS,Drischler:2020yad_matter}. For this, the expansion coefficients are assumed to follow a joint Gaussian distribution
\begin{equation}
    c_i(\mathbf{y})\ \vert\ \theta  \sim \mathcal{GP}\left[\mu,\kappa (\mathbf{y},\mathbf{y}^\prime; \theta)\right] \, ,
    \label{eq:cGP}
\end{equation}
with the mean $\mu$ and the covariance $\kappa(\mathbf{y},\mathbf{y}^\prime; \theta)$. For the following we are agnostic regarding the sign of the coefficients and hence set the mean to zero, $\mu = 0$. $\theta$ denotes a set of hyperparameters that characterizes the covariance. The covariance accounts for correlations of the coefficients at different densities and proton fractions. For its parametrization, we choose the common representation of a radial basis function (RBF) kernel
\begin{align}
    \kappa(\mathbf{y},\mathbf{y}^\prime;\theta = \{\bar{c}^2, \mathbf{L}\}) &= \bar{c}^2 \exp \left[-\frac{1}{2}(\mathbf{y}-\mathbf{y}^\prime)^{\text{T}} \mathbf{L} (\mathbf{y}-\mathbf{y}^\prime)\right] \, , \\
    &= \bar{c}^2 r(\mathbf{y}, \mathbf{y}^\prime; \mathbf{L}) \, .  
\end{align}
This choice reflects the smooth and stationary behavior of the coefficient functions. Stationarity, in this context, implies the underlying assumption that the coefficient functions behave similarly across the whole input space. The distribution is characterized by the marginal variance $\bar{c}^2$, describing the size of variations about the mean, and the diagonal correlation matrix $\mathbf{L}$, containing the correlation lengths $\ell_{n}$ and $\ell_{x}$ along the independent variables:
\begin{equation}
\mathbf{L} = 
\begin{pmatrix}
\ell_{n}^{-2} & 0 \\
0 & \ell_{x}^{-2} \\
\end{pmatrix} \, .
\end{equation}

With the expansion coefficients following a multivariate normal, the EFT truncation errors $\delta E_k$ follow directly from the fundamental property of Gaussian random variables being closed under addition and matrix multiplication. Together, Eqs.~\eqref{eq:DeltaEk} and~\eqref{eq:cGP} yield that the EFT truncation error $\delta E_k$ is itself a Gaussian random variable~\cite{Furnstahl:2015rha,Drischler:2020hwi_EOS,Drischler:2020yad_matter}, such that 
\begin{equation}
    \delta E_k\ \vert\ \theta \sim \mathcal{GP} \left[ 0, \bar{c}^2 R_{\delta k}(\mathbf{y},\mathbf{y}^\prime;\mathbf{L})\right] \, .
    \label{eq:truncationGP}
\end{equation}
With the geometric sum in the expansion parameter $Q(\mathbf{y})$ in Eq.~\eqref{eq:DeltaEk}, the covariance of this GP is given by
\begin{equation}
    R_{\delta k}(\mathbf{y},\mathbf{y}^\prime;\mathbf{L}) = E_{\text{ref}}(\mathbf{y}) E_{\text{ref}}(\mathbf{y}^\prime)\frac{\left[Q(\mathbf{y})Q(\mathbf{y}^\prime)\right]^{k+1}}{1-Q(\mathbf{y})Q(\mathbf{y}^\prime)} r(\mathbf{y},\mathbf{y}^\prime; \mathbf{L}) \, .
    \label{eq:uckernel}
\end{equation}

With $Q$ being fixed, the only way for information to propagate from the known order-by-order predictions $E_k$ to the unknown EFT orders and thus the EFT uncertainty $\delta E_k$ is by optimizing the hyperparameters $\theta$. In particular, the hyperparameters $\bar{c}$, $\ell_{n}$, and $\ell_{x}$ are adjusted such that random samples of the GP for the coefficients match the distribution of the calculated ones. The quantification of this match is given by the likelihood distribution that emerges from the calculated coefficients being described by a GP with the kernel as covariance, and we adopt the hyperparameters that maximize the likelihood.
Additionally we consider prior information on the order-by-order convergence. The details of the training and prior are discussed below.

\subsection{Training and optimization choices}

\begin{figure*}[t!]
    \centering
    \includegraphics[width=\linewidth]{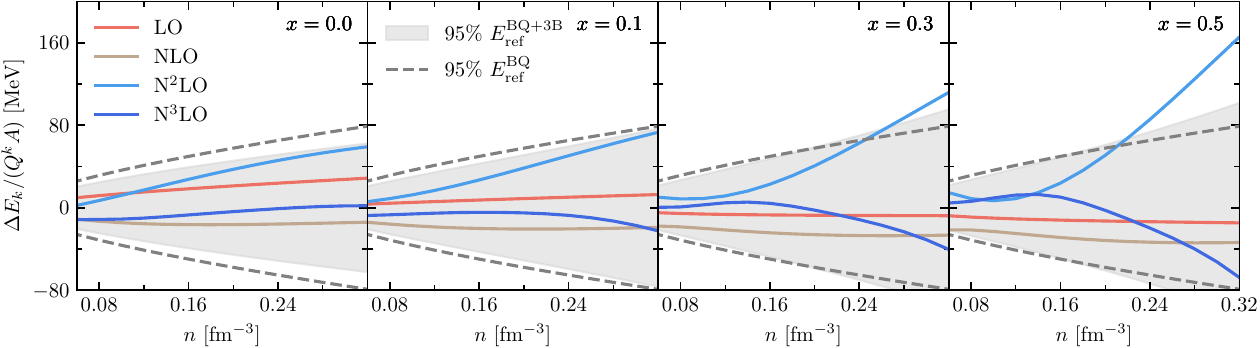}
    \caption{Corrections to the energy per particle from LO to N$^3$LO divided by the respective powers of the expansion parameter $Q$ as function of density $n$ at different proton fractions from PNM ($x = 0$, left) to SNM ($x = 0.5$, right). The 95\% credibility ranges of the EFT uncertainties from the GP are shown for the reference energies $\erefB$ (gray region) and $\erefBQ$ (dashed gray lines).}
    \label{fig:norm_res}
\end{figure*}

In the previous section, we argued the stationarity of the expansion coefficients, justifying the use of the RBF kernel and leading to the EFT uncertainty estimates. Still, choices that can be made in the expansion influence the behavior of the expansion coefficients. In Fig.~\ref{fig:norm_res}, we investigate the order-by-order corrections to the energy per particle divided by the respective powers of the expansion parameter $Q$,
\begin{equation}
    \frac{\Delta E_k}{Q^k} = \frac{E_k(n,x)-E_{k-1}(n,x)}{Q^k(n,x)} = \eref(n,x) c_k(n,x) \,,
\label{eq:normresiduals}
\end{equation}
before adopting a reference energy. In Fig.~\ref{fig:norm_res},  the impact of the 3N contributions entering at N$^2$LO becomes evident at larger densities and proton fractions. Beyond $x>0$, the N$^2$LO and N$^3$LO energy corrections increase in magnitude and show additional curvature. With the same argument as before, the energy corrections~\eqref{eq:normresiduals} also follow a GP, the covariance of which again depends on the reference energy $\eref$. With the optimized set of hyperparameters (the details of the construction are discussed below), it is then straightforward to construct credibility ranges from the GP. In Fig.~\ref{fig:norm_res}, we show the resulting 95\% credibility ranges for the reference energy $\erefBQ$ from Eq.~\eqref{eq:ref} as dashed gray lines. By construction, these GP bands cannot capture the varying behavior in proton fraction. In particular, the uncertainty for PNM appears to be overestimated, while the uncertainty for proton fractions near SNM is underestimated.

Based on these arguments, we promote the use of a proton-fraction-dependent reference energy $\erefB$. In addition to the LO-informed $\erefBQ$, we include a contribution motivated by 3N forces, which become more important at high densities and are stronger in SNM. To this end, we adapt the reference energy as
\begin{equation}
    \frac{\erefB (n,x)}{A}  =  \frac{\erefBQ (n)}{A} + \SI{4}{MeV} \left( n/n_0 \right)^{6/3} (1-\delta^2) \,,
\label{eq:newrefenergy}
\end{equation}
with the neutron excess $\delta = (1 - 2x)$. The exponent of the second term is chosen to mimic 3N contributions scaling approximately with $n^3$ in the energy density, while the prefactor can be motivated by the strength of $c_D$ and $c_E$ contributions to the 3N energy (see, e.g., Ref.~\cite{Report_Kai}). Using this $x$-dependent reference energy, we construct a new GP (details again below). The corresponding 95\% credibility ranges are given by the gray region in Fig.~\ref{fig:norm_res}. While being slightly more strict at low proton fractions, the additional term in the reference energy $\erefB$ leads to a widening of the bands at larger proton fractions and higher densities, leaving the N$^2$LO results less outlying compared to the bands obtained with $\erefBQ$. For completeness, the expansion coefficients extracted using the proton-fraction-dependent reference energy $\erefB$ are given in the Appendix. The additional 3N-inspired term leads to better behaved expansion coefficient at larger densities and $x>0$. In particular, the coefficients now share a more common scale across all $n$ and $x$ considered.

The training of the GP is performed based on the calculated expansion coefficients $c_i$ up to N$^3$LO. We do not consider the LO coefficient $c_0$ as it informs the choice of the reference energy. To improve stationarity along the independent variables, we work with the momentum scale $k_{\text{f}} = (3\pi^2n)^{1/3}$ as variable instead of the density. Note that $k_{\text{f}}$ only coincides with a Fermi momentum for PNM.

\begin{figure}[t!]
    \centering
    \includegraphics[width=\linewidth]{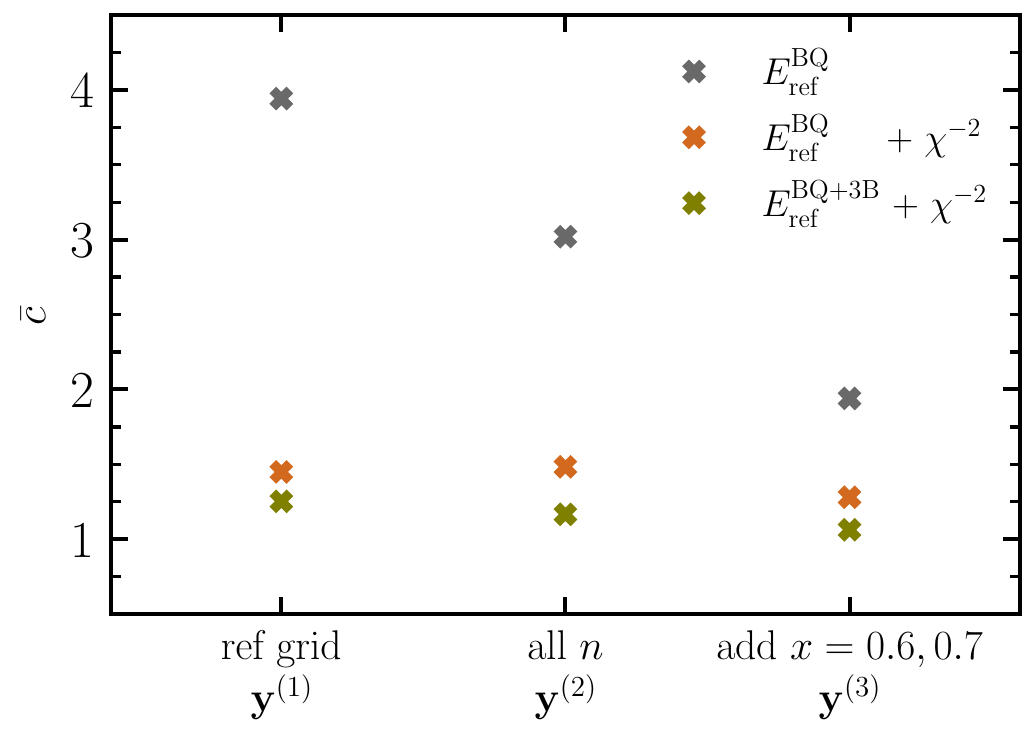}
    \caption{Optimized marginal standard deviation ${\bar c}$ for the three training sets (see text for details), the two different reference energies $\erefBQ$ and $\erefB$, and optimization choices (inverse-$\chi^2$ prior or uniform prior, when not stated).}
    \label{fig:sensitivity}
\end{figure}

Generally, the choice of training set has an impact on the values of the extracted hyperparameters. Figure~\ref{fig:sensitivity} shows the value of the hyperparameter $\bar{c}$ for different training sets. Our reference training set $\mathbf{y}^{(1)} = (\mathbb{N}^{(1)}, \mathbb{X}^{(1)})$ includes the following combinations of densities and proton fractions:
\begin{align}
    \mathbb{N}^{(1)} &= [0.06, 0.08, 0.12, 0.16, 0.2, 0.24, 0.28, 0.32]\fmiq \, , \\
    \mathbb{X}^{(1)} &= [0, 0.1, 0.2, 0.3, 0.4, 0.5] \, .
\end{align}
For comparison we show results for a second training set $\mathbf{y}^{(2)} = (\mathbb{N}^{(2)}, \mathbb{X}^{(1)})$ that contains the complete set of density grid points, i.e.,
\begin{align}
    \mathbb{N}^{(2)} &= [0.04, 0.06, ..., 0.30, 0.32] \fmiq \, ,
\end{align}
and a third training set $\mathbf{y}^{(3)} = (\mathbb{N}^{(1)}, \mathbb{X}^{(2)})$ that instead contains results at additional proton fractions
\begin{align}
    \mathbb{X}^{(2)} &= [0.0, 0.1, 0.2, 0.3, 0.4, 0.5, 0.6, 0.7] \, .
\end{align}

Instead of a uniform prior on the marginal variance, we encapsulate the convergence behavior with powers of $Q$ by implementing the expectation of the naturalness of the expansion coefficients as prior information quantified by an inverse-$\chi^2$ distribution
\begin{equation}
    pr(\bar{c}^2) = \frac{(\nu_0\tau_0^2/2)^{\nu_0/2}}{\Gamma(\nu_0/2)(\bar{c}^2)^{1 + \nu_0/2}} \exp\left(-\frac{\nu_0\tau_0^2}{2\bar{c}^2}\right) \, ,
\end{equation}
setting the expectation value to $\mathbb{E} = \nu_0\tau^2_0/(\nu_0 - 2) = 1$. For the certainty of our prior guess for $\bar{c}^2$, we choose $\nu_0 = 10$ as a variance parameter, and set the scaling parameter $\tau_0$ accordingly (see Refs.~\cite{Drischler:2020hwi_EOS,Drischler:2020yad_matter}). In particular, this prior is conjugate: Updating the likelihood of the hyperparameter to match the data using Bayes' theorem results in a posterior of the same functional form, with updated parameters $\nu$ and $\tau$. As there are no conjugate priors for the correlation lengths and we do not have specific knowledge about their size, we choose a uniform prior on $\ell_n$ and $\ell_x$. Marginalizing the likelihood over $\bar{c}^2$, the correlation lengths are optimized using the maximum value of the remaining likelihood. Then, $\bar{c}$ is updated analytically.

As shown in Fig.~\ref{fig:sensitivity}, the sensitivity of the optimized marginal standard deviation $\bar{c}$ to the employed training set is small when the inverse-$\chi^2$ prior is used. However, as discussed above, including results at proton fractions $x>0.5$ has a significant impact on the correlation length $\ell_{x}$ due to isospin symmetry around $x=0.5$. Since we are mainly interested in an accurate description of neutron-rich matter, we exclude in the following the proton fractions $x = 0.6$ and $0.7$ from the training set to avoid an artificial bias. 

\subsection{Model-checking diagnostics}

We need to verify that the constructed GP, with the discussed choices and assumptions, reasonably describes the underlying system. To this end, we employ model-checking diagnostics. In particular, we evaluate our GP model with the optimized set of hyperparameters $\theta$ at a set of validation points~$\mathbf{y}_{\text{val}}$, which does not overlap with the set of training points~$\mathbf{y}^{(1)}$. We follow the model-checking diagnostics from Ref.~\cite{Melendez:2019izc_EFT}. For a more detailed discussion as well as a broader range of diagnostics for GPs, we refer the reader to Ref.~\cite{Bastos}.

\begin{figure*}[t!]
    \centering
    \includegraphics[width=\linewidth]{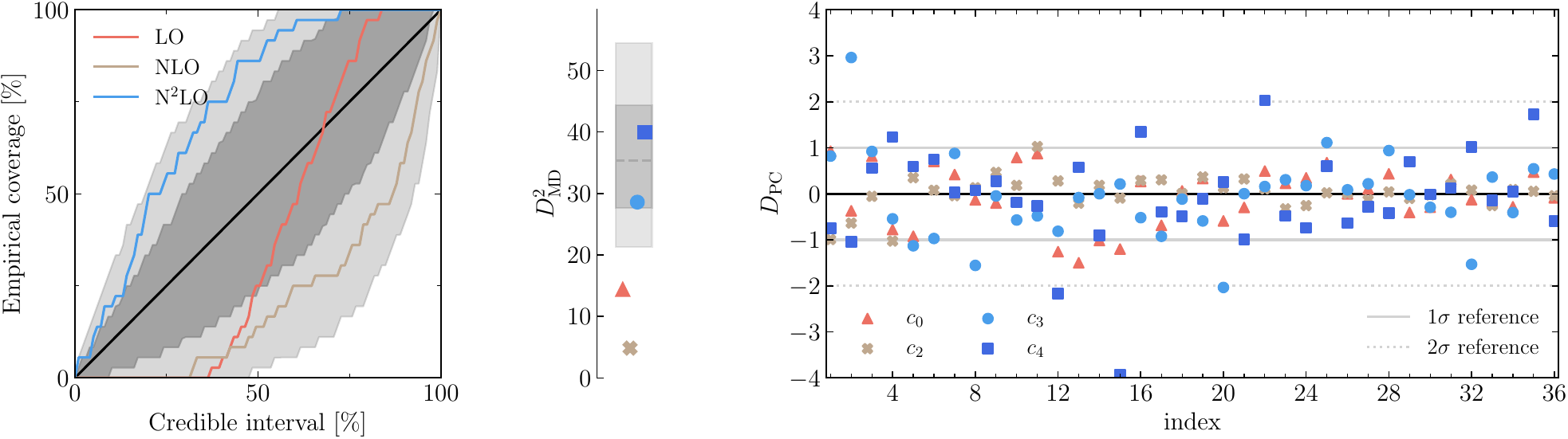}
    \caption{Model-checking diagnostics for the GP constructed with the training set $\mathbf{y}^{(1)}$, the reference energy $\erefB$, and the inverse-$\chi^2$ prior. Left: credibility interval diagnostics. Middle: Mahalanobis distance.
    Right: Pivoted Cholesky errors. For each diagnostics, the dark (light) gray areas/lines give 68\% (95\%) credibility intervals of the corresponding reference distributions. See text for details.}
    \label{fig:diagnostics}
\end{figure*}

A commonly employed diagnostic to assess the size of EFT uncertainties is the credibility interval diagnostic. The assumption is that a well-behaved 100$\alpha$\% credibility interval contains approximately 100$\alpha$\% of validation points. For a given $\alpha  \in [0,1]$ and a set of $M$ validation points $f_i \in \mathbf{f}_{\text{val}} = \mathbf{f}(\mathbf{y}_{\text{val}})$, one constructs credibility intervals $\text{CI}_i(\alpha)$ and  computes the empirical coverage, i.e., the fraction of points included by these:
\begin{equation}
    D_{\text{CI}}(\alpha, \mathbf{f}_{\text{val}}) = \frac{1}{M} \sum_{i=1}^{M} \boldsymbol{1}[f_i \in \text{CI}_i(\alpha)] \, .
    \label{eq:credibleintervaldiagnostic}
\end{equation}
Here, $\boldsymbol{1}$ is an indicator function, returning either $1$ or $0$. Intuitively, for the case of order-by-order convergence, the credibility intervals are given by the EFT uncertainties at order $k$, and we test whether they include the prediction of the upcoming order $k+1$. To compare the results to a meaningful reference distribution, instead of the energy per particle we consider the norm residuals from Eq.~\eqref{eq:normresiduals}. It is then straightforward to draw samples of the underlying GP and construct credibility bands from the resulting distribution of the empirical coverage.

For a meaningful graphical evaluation, one should perform this computation for a set of $\alpha \in [0,1]$. Curves lying above the diagonal then imply overly conservative uncertainties while curves lying below the diagonal imply overly strict uncertainties. We show the results for our optimized GP in the left panel of Fig.~\ref{fig:diagnostics}. All orders lie within the $95\%$ credibility intervals of the reference, indicating well constructed truncation errors. The behavior of the different orders is, however, mostly determined by the convergence of the chiral EFT expansion itself and less by the specifics of our model. The 3N contributions entering at N$^2$LO cannot be captured by previous orders, resulting in too strict uncertainties at LO. Similarly, the N$^3$LO NN and 3N corrections are small, resulting in too conservative uncertainties at  N$^2$LO.

As a generalization of the sum of squared residuals, the Mahalanobis distance includes correlations among variables to assess the correctness of emulator predictions. For validation data $\mathbf{g}_{\text{val}}$, mean $\boldsymbol{\mu}$, and correlation matrix $K$ of a GP it is given by
\begin{equation}
    D_{\text{MD}}^2 = (\mathbf{g}_{\text{val}}-\boldsymbol{\mu})^{\text{T}} K^{-1} (\mathbf{g}_{\text{val}}-\boldsymbol{\mu}) \, .
\end{equation}
Specifically, we want to test whether the coefficients $c_i$ are realistic realizations of the trained GP, i.e., whether the assumption of Eq.~\eqref{eq:cGP} is valid. To this end, for each order $k$, we compare the coefficients $c_k(\textbf{y}_{\text{val}}) = \mathbf{g}_{\text{val}}$ against the optimized GP with $\boldsymbol{\mu} = 0$ and $K=\kappa(\mathbf{y}_{\text{val}},\mathbf{y}_{\text{val}};\theta)$. Very large or very small values of $D_{\text{MD}}^2$ can then identify possible outliers and hint at either poorly optimized hyperparameters or an issue with the choice of model. For comparison, we need a reference distribution. With the underlying process being Gaussian, it is straightforward to show that the Mahalanobis distance follows a $\chi^2$ distribution with $M$ degrees of freedom, i.e., $D_{\text{MD}}^2 \sim \chi^2_M$. The middle panel of Fig.~\ref{fig:diagnostics} shows the results for our optimized GP. The dashed line marks the median and the dark gray and light gray areas the 68\% and 95\% credibility intervals of the reference distribution. For $c_3$ and $c_4$, the $D_{\text{MD}}^2$ values are reasonably sized, while the result for $c_2$ lies significantly below the reference distribution. This is expected though, because $c_2$ is small and shows only minimal variation over the whole $\mathbf{y}$-plane (see Figs.~\ref{fig:c_n},~\ref{fig:c_x}, and~\ref{fig:norm_res}), resulting in a small $D_{\text{MD}}^2$ value. Hence, $c_2$ may be treated as an outlier in this regard.

A more in-depth analysis can be made by performing a variance decomposition on the Mahalanobis distance. 
By writing $K = GG^T$, we can introduce the vector 
\begin{equation}
    D_G = G^{-1}(\mathbf{g}_{\text{val}}-\boldsymbol{\mu}) \, .
    \label{eq:variancedecomposition}
\end{equation}
As there are multiple ways to decomposing the kernel matrix, this diagnostic is not unique. A decomposition which has proven to be particularly useful is the pivoted Cholesky decomposition, which constructs the unique lower triangular matrix $L$, such that $K = L^TL$, and introduces an additional permutation of the validation data. This permutation is based on maximizing the diagonal elements of the decomposition matrix, i.e., they are sorted according to their conditional variance regarding the previous element. With Eq.~\eqref{eq:variancedecomposition}, this produces a decomposition vector $D_{\text{PC}}$, which is independent of the order of validation points and allows for an evaluation of the correctness of variance and length scale: Unusually many small or large entries of $D_{\text{PC}}$ imply a wrongly estimated variance, and a convergence or divergence of entries at large indices may point to a wrongly calibrated correlation length. We show the computed $D_{\text{PC}}$ vector plotted versus the index of its entries in the right panel of Fig.~\ref{fig:diagnostics}. Ideally, one would expect the points to be standard Gaussian distributed and not show any pattern along the index. Approximately, this is the case. However, for the same reason as before, the variation of the $c_2$ results around the mean is notably small.  Although $c_0$ is excluded from our optimization, we show its diagnostics for completeness. Due to its strong relation to the reference energy, the resulting $D_{\text{MD}}^2$ and $D_{\text{PC}}$ values are small as well.

\section{EOS results with EFT uncertainties}
\label{sec:results}

With Eq.~\eqref{eq:truncationGP} and the optimized set of hyperparameters, we can construct EFT uncertainty bands for the energy per particle, i.e., estimate $\delta E_k$. In addition, we need to evaluate the chiral EFT result for the energy per particle at this order, $E_k$, for arbitrary density and proton fraction. As in Ref.~\cite{Keller:2022crb}, we construct a GP emulator for this. We again employ a RBF kernel and choose the mean of the GP to be zero. Since we expect the computed energies to contain solver uncertainties from the Monte Carlo integration, as in Ref.~\cite{Keller:2022crb} we include a Gaussian noise term on the energy per particle with $\sigma_n \approx20 \, \text{keV}$ in the GP prior. We treat the free Fermi gas (FG) contribution analytically and use the GP to emulate only the interaction part $E_{\text{int}}/A = E/A - E_{\text{FG}}/A$. This was found to deliver a more stable behavior of the GP and its derivatives at small densities \cite{Keller:2022crb}. We also found that defining a constant noise term $\sigma_n$ leads to unphysical variations at low densities, where energies are small and data points are close to each other. This issue can be resolved, however, by transforming the input space such that the low density points are further stretched out; the GP can then emulate these data points with smaller deviations. As before, we use the momentum scale $k_{\text{f}}$ instead of the density as the input space for our GP, in addition to the proton fraction. To reduce numerical artifacts, we also normalized the input variables to $[0,1]$. The hyperparameters are then extracted using uniform priors and a maximum a posteriori estimate. Eventually, we obtain $E_k$ as the mean of the GP posterior conditioned upon the energy per particle data of order $k$. For each order, a separate GP is constructed.

\subsection{Nuclear matter at arbitrary proton fractions}

\begin{figure}[t!]
    \centering
    \includegraphics[width=\linewidth]{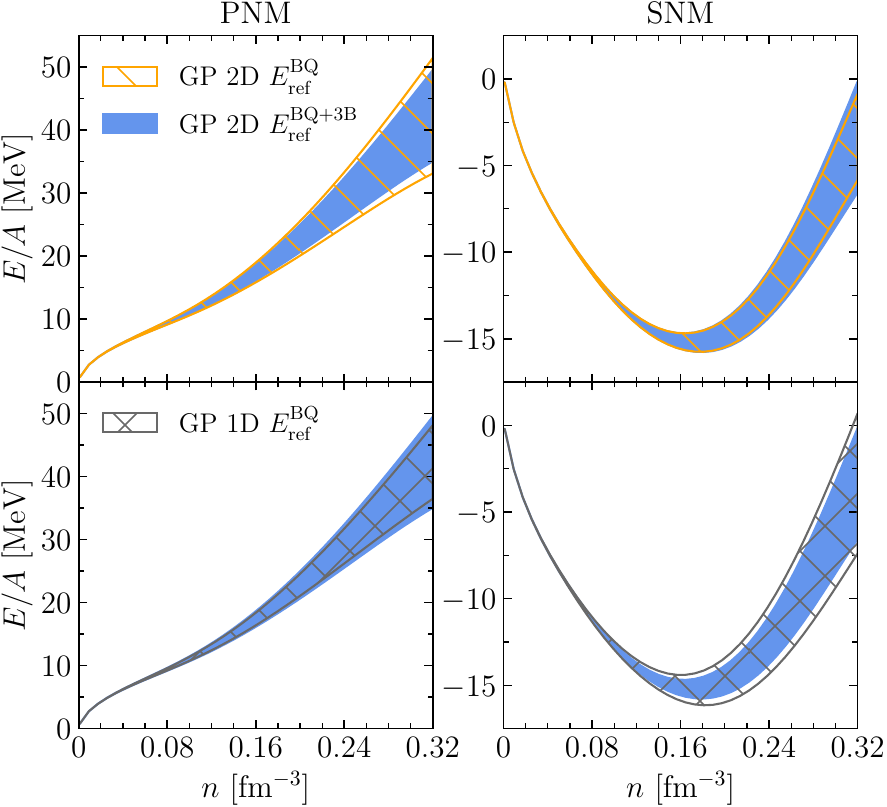}
    \caption{Energy per particle at N$^3$LO as function of density for PNM (left panels) and SNM (right panels). The top panels compare the 68\% credibility ranges for the EFT uncertainties obtained by training the GP on  asymmetric matter results with the reference energy $\erefB$ and $\erefBQ$, respectively. The bottom panels compare the two-dimensional GP with $\erefB$ to results obtained from a one-dimensional GP with $\erefBQ$ for PNM or SNM separately.}
    \label{fig:compare_E}
\end{figure}

After the optimization, the evaluation of the GP emulator at every given point in density and proton fraction is quick. The energy per particle can thus be easily evaluated including the EFT truncation uncertainty. In Fig.~\ref{fig:compare_E} we show the 68\% credibility ranges for the EFT uncertainties of the energy per particle as a function of density for PNM and SNM. The blue bands show the results obtained with the new $x$-dependent reference energy $\erefB$ from Eq.~\eqref{eq:newrefenergy}.
In the top panels we compare those results to the bands extracted from an analogous GP with the $x$-independent reference energy $\erefBQ$ from Eq.~\eqref{eq:ref}. The bottom panels compare the new two-dimensional (2D) GP to the 68\% credibility ranges for 1D GPs trained for PNM or SNM separately.

We observe that the EFT uncertainties in PNM are slightly wider for the 2D GPs with both reference energies compared to the 1D GP, and vice versa for SNM. Since the credibility ranges for the energy per particle scale directly with the marginal variance $\bar{c}^2$, this is intuitive: While the 2D GP is extended by an additional hyperparameter $\ell_x$ to account for correlations along the $x$-dimension, in our stationary kernel the marginal variance is constant along both independent variables. However, the overall scale varies along~$x$ so that this cannot be captured with the $x$-independent reference energy $\erefBQ$. As can be seen in Figs.~\ref{fig:c_n} and~\ref{fig:c_x}, the magnitude of the expansion coefficients in general increases when moving from $x=0$ to $x=0.5$, leading to a significantly greater $\bar{c}$ for the SNM 1D GP compared to the PNM 1D GP. As a result, going to the 2D training, the PNM bands will be inflated while the SNM bands will be compressed. Including the $x$-dependence in the reference energy counteracts this effect. The additional term motivated by 3N contributions leads to a decrease in magnitude of the expansion coefficients at large densities and $x>0$, such that the coefficients share a more common scale across all $x$. As a result, the 2D GP uncertainties with the new reference energy $\erefB$ are more similar to the 1D GP uncertainties in both PNM and SNM, indicating that the new 2D GP can better reproduce the EFT uncertainties for all $x$.

\begin{figure}[t!]
    \centering
    \includegraphics[width=\linewidth]{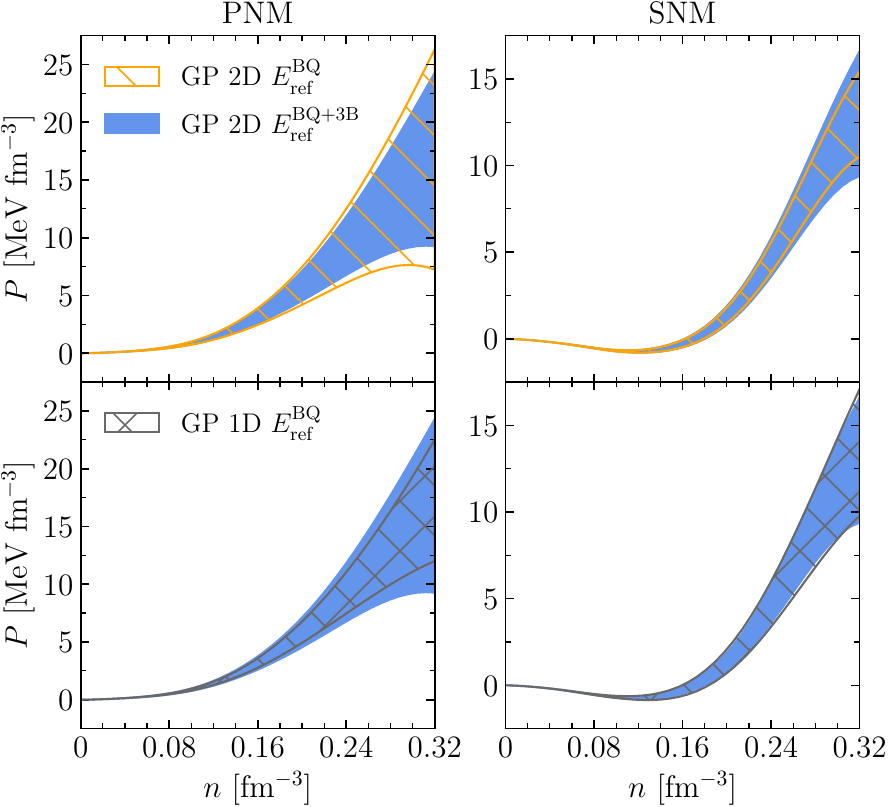}
    \caption{Same as Fig.~\ref{fig:compare_E} but for the pressure as a function of density.}
    \label{fig:compare_P}
\end{figure}

\begin{figure}[t!]
    \centering
    \includegraphics[width=0.7\linewidth]{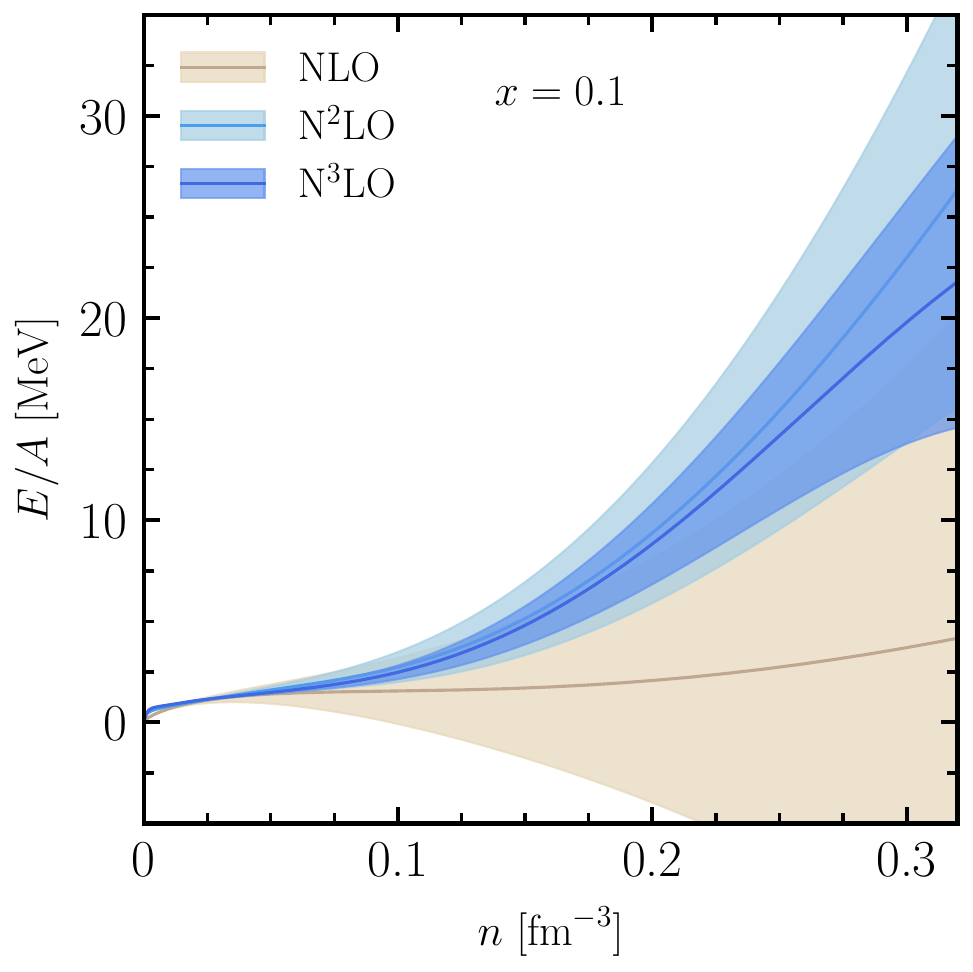}
    \caption{Energy per particle from NLO to N$^3$LO as function of density at proton fraction $x=0.1$. The bands show 68\% credibility ranges for the EFT uncertainties from the 2D GP with reference energy $\erefB$.}
    \label{fig:Ex10}
\end{figure}

A particular strength of representing the EOS by GPs for the mean and the EFT uncertainties lies in the ability to calculate derivatives analytically. This is a direct consequence of multivariate Gaussians, so that GPs remain closed under linear transformations. From the 2D GP emulator for the energy per particle it is thus straightforward to compute the pressure 
\begin{equation}
    P = n^2 \frac{\partial(E/A)}{\partial n} \, .
    \label{eq:pressure}
\end{equation}
\begin{figure*}[t!]
    \centering
    \includegraphics[width=0.9\linewidth]{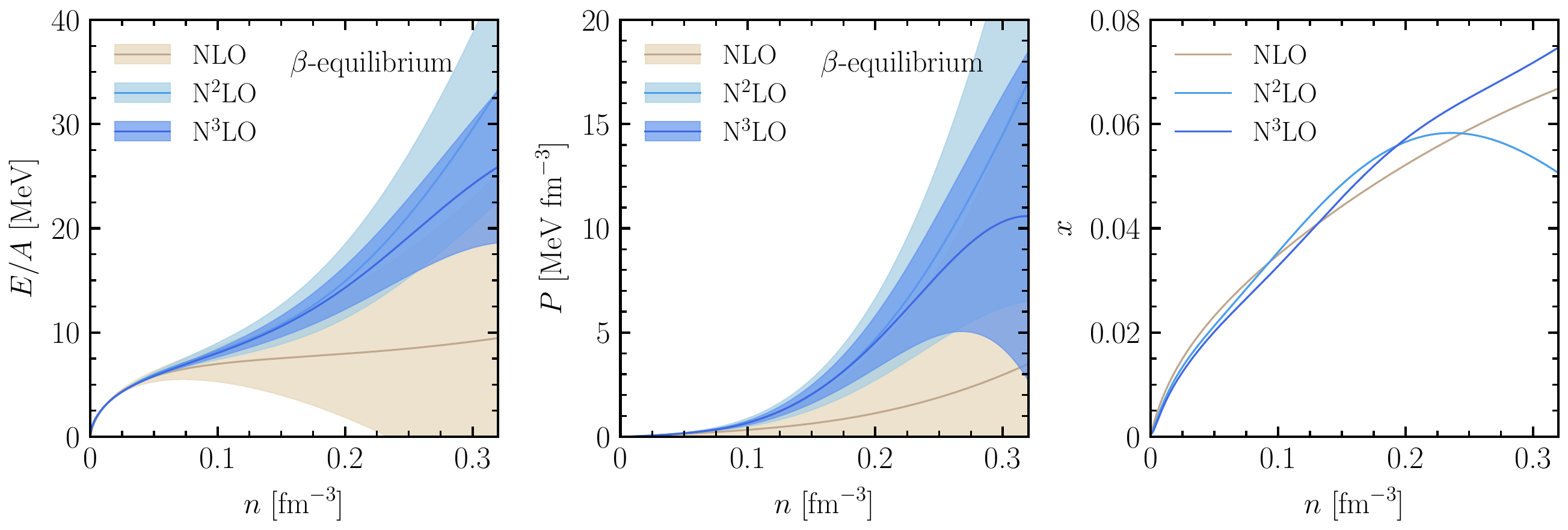}
    \caption{Energy per particle (left panel), pressure (middle panel), and proton fraction (right panel) for matter in $\beta$ equilibrium from NLO to N$^3$LO as function of density. The bands show 68\% credibility ranges for the EFT uncertainties obtained from the 2D GP for the energy per particle with reference energy $\erefB$.}
    \label{fig:beta}
\end{figure*}
Due to the series definition of Eq.~\eqref{eq:seriesdefinition}, this translates directly to the EFT uncertainties as well. Analogous to the energy per particle, we show the 68\% credibility ranges for the EFT uncertainties of the pressure as a function of density for PNM and SNM in Fig.~\ref{fig:compare_P}. We find similar comparisons for the pressure as for the energy per particle. In the following, we will therefore use the 2D GP with the new $x$-dependent reference energy $\erefB$.

With the 2D GP, the EOS is not restricted to PNM and SNM, but can be evaluated at arbitrary proton fraction. As an example, we show in in Fig.~\ref{fig:Ex10} the order-by-order EFT uncertainties for the energy per particle as a function of density at proton fraction $x=0.1$. We find systematic EFT uncertainty bands from NLO to N$^3$LO, with no indication of the breakdown of chiral EFT up to the largest density considered $2 n_0$.

Finally, we remark that for our EFT uncertainty kernel, Eq.~\eqref{eq:uckernel}, additional care would need to be taken when computing derivatives for $x=0.5$. This is because the expansion parameter $Q$ is defined by a maximum function, so that additional smoothening would be needed when taking the derivative in $x$-direction at the point where the proton and neutron densities match. We leave this technical aspect to future work, as this paper is focused on neutron-rich conditions.

\subsection{Matter in $\beta$ equilibrium}

Next, we explore neutron star matter which is in $\beta$ equilibrium of neutrons undergoing $\beta$ decays and protons capturing electrons. The chemical potential $\mu_n$ of neutrons then equals the sum of the proton chemical potential $\mu_p$ and electron chemical potential $\mu_e$, assuming the (anti-)neutrinos decouple from the neutron star so that their chemical potentials are zero. Measuring the chemical potentials from the rest masses, the $\beta$ equilibrium condition thus read:
\begin{equation}
    \mu_n + m_n = \mu_p + m_p + \mu_e +m_e \, .
\label{eq:betaequilibrium}
\end{equation}
The electrons are taken to be non-interacting so that their  chemical potential is
\begin{equation}
    \mu_e + m_e = \left[\left(3\pi^2 n_e\right)^{2/3}+m_e^2\right]^{1/2} \, ,
\label{eq:electronchemicalpotential}
\end{equation}
with the electron density determined by charge neutrality $n_e = n_p$. The chemical potential of the nucleons ($\tau=n,p$) is derived from the energy per particle as
\begin{align}
    \mu_{\tau} &= \frac{\partial (E/V)}{\partial n_\tau} = \frac{\partial (E/A \cdot n)}{\partial n_\tau} \\
    &= E/A + n \cdot \left[\frac{\partial E/A}{\partial n}\frac{\partial n}{\partial n_\tau} +\frac{\partial E/A}{\partial x}\frac{\partial x}{\partial n_\tau}\right] \, ,
\label{eq:nucleonchemicalpotentials}
\end{align}
where $\partial n / \partial n_\tau = 1$  and $\partial x / \partial n_n = -x/n$ for neutrons and $\partial x / \partial n_p = (1-x)/n$ for protons, respectively.

\begin{figure}[t!]
    \centering
    \includegraphics[width=\linewidth]{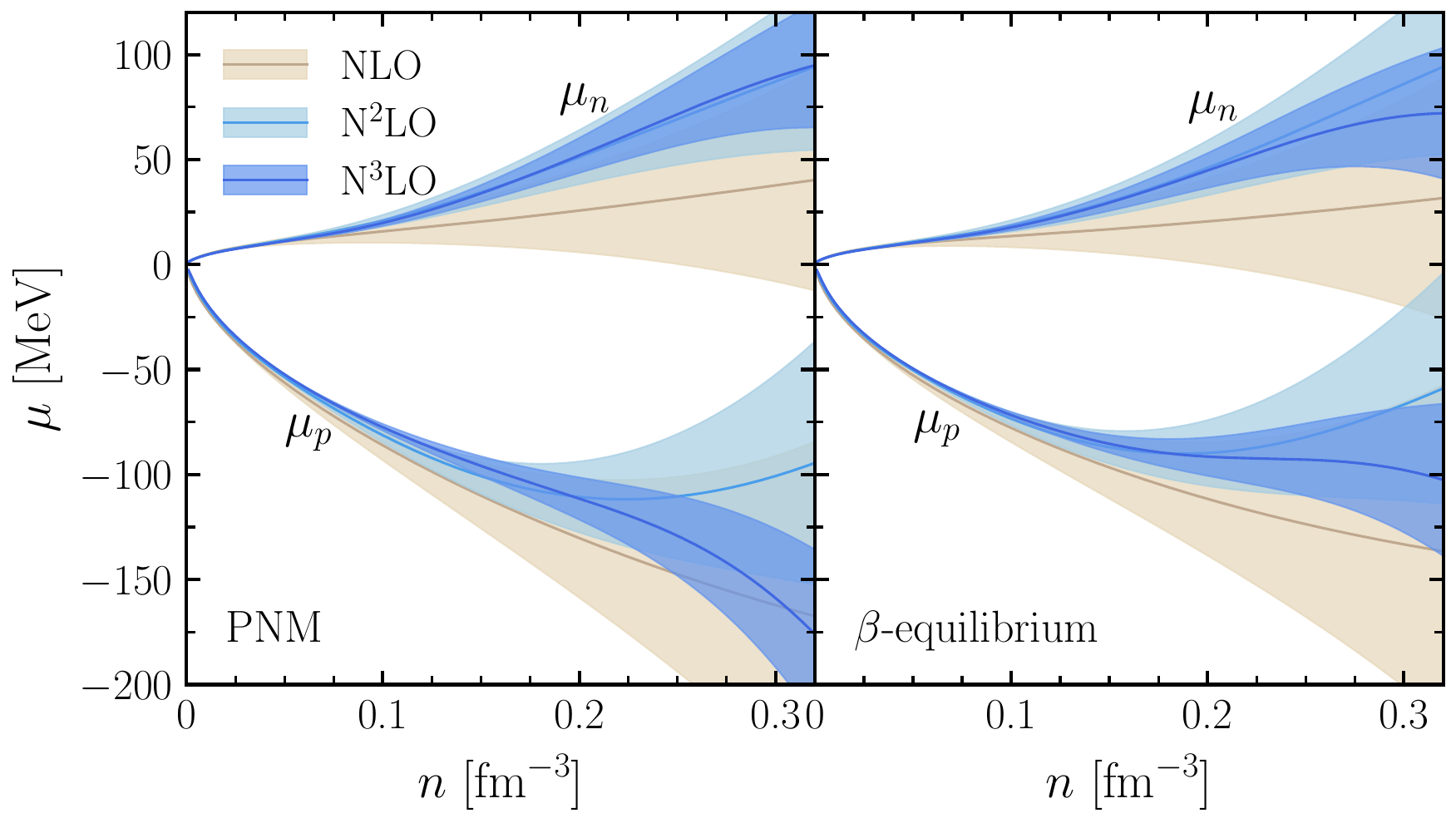}
    \caption{Chemical potentials of neutrons $\mu_n$ and protons $\mu_n$ for PNM (left panel) and matter in $\beta$ equilibrium (right panel) from NLO to N$^3$LO as function of density. The bands show 68\% credibility ranges for the EFT uncertainties obtained from the 2D GP for the energy per particle with reference energy $\erefB$.}
    \label{fig:mu}
\end{figure}

We calculate the chemical potentials of the nucleons from the 2D GP and determine the proton fraction in $\beta$ equilibrium. We stress that this does not require phenomenological parametrizations to interpolate between PNM and SNM. The resulting proton fraction in $\beta$ equilibrium as a function of density is shown in the right panel of Fig.~\ref{fig:beta}. At all orders from NLO to N$^3$LO the proton fraction is very similar up to around $1.5 n_0$. For N$^3$LO, the proton fraction in uniform matter increases monotonously until around $7.5\%$ for $2 n_0$. Using the proton fraction in $\beta$ equilibrium, we show the energy per particle and the pressure including EFT uncertainties in the left and middle panels of Fig.~\ref{fig:beta}. The order-by-order behavior is very systematic up to $2 n_0$ for the energy and $1.5 n_0$ for the pressure. For the higher densities $n > 1.5 n_0$ the N$^3$LO energy increases less strongly than at N$^2$LO leading to a decrease of the pressure towards $2 n_0$. This becomes more pronounced for the pressure due to the derivative involved. Whether this is due to regulator effects or from coming closer to the breakdown scale needs further investigation.

Finally, the corresponding neutron and proton chemical potentials for $\beta$ equilibrium, as well as for PNM, are shown in Fig.~\ref{fig:mu}. Because the effects on the proton chemical potential in dense neutron-rich matter are larger, the order-by-order behavior for $\mu_p$ and the EFT uncertainties show more of a deviating trend for densities $n > 1.5 n_0$.

\section{Constructing consistent crust equations of state}
\label{sec:crust}

In the outmost part of the neutron star, neutrons and protons are bound in neutron-rich nuclei, forming a lattice within a sea of electrons. This outer crust, which can be probed in terrestrial experiments, is better understood and constrained; a commonly employed EOS was constructed by Baym, Pethick, and Sutherland (BPS)~\cite{BPS}. Deeper in the star, the density and pressure rise and nuclei become increasingly neutron-rich. At some density, neutrons start to drip out of the nuclei and form a neutron fluid. This neutron drip phase marks the beginning of the inner crust. At the bottom of the crust, also protons may drip out of the nuclei, resulting in in an equilibrium of two phases which both contain neutrons and protons. The existence of proton drip was recently established in the uniform phase based on chiral EFT calculations and general considerations~\cite{Keller_2024}. Here, we show that it also persists including surface and Coulomb corrections. Near the crust-core boundary, nuclei (or high-density clusters) may undergo various transitions from spherical shape to, e.g., rod-like or plate-like structures. The realizations of these nuclear pasta phases and their astrophysical implications are an open area of research.

In this section, we construct neutron star EOSs for the inner crust including surface and Coulomb corrections, which are consistent with the chiral EFT results and uncertainties for uniform matter from Sec.~\ref{sec:results}. We focus on results at N$^3$LO.

\subsection{Inner crust from chiral EFT results}
\label{sec:crusttheory}

To determine the inner crust EOS, we employ a compressible liquid drop model within the Wigner-Seitz approximation. For neutron drip, we assume a spherical Wigner-Seitz cell of Volume $V_{\text{W}}$, containing PNM (phase $1$). In its center, the nucleus is modeled as a sphere of asymmetric nuclear matter, with Volume $V_0$ and proton number $Z$ (phase $2$). In both phase 1 and phase 2, we assume the nucleon densities $n_n$ and $n_p$ to be uniform. Additionally, the Wigner-Seitz cell contains a  uniformly distributed relativistic electron gas with electron density $n_{e}$.

For a stable phase coexistence, the highest occupied energy state of a neutron must be identical in both phases. This condition is fulfilled if the neutron chemical potentials match. Equilibrium additionally requires the pressure in both phases to be equal, i.e., the conditions for neutron drip read
\begin{align}
    P^{(1)} &= P^{(2)} \, , \\
    \mu_n^{(1)} &= \mu_n^{(2)} \, , \\
    \mu_p^{(1)} &> \mu_p^{(2)} \, .
\label{eq:neutrondrip}
\end{align}
The latter condition guarantees that no protons drip from phase 2 into phase 1. Since neutron star matter is in $\beta$ equilibrium, Eq.~\eqref{eq:betaequilibrium} provides an additional constraint. 

The pressure and nucleon chemical potentials follow directly from the energy per particle [see Eqs.~\eqref{eq:pressure} and~\eqref{eq:nucleonchemicalpotentials}].
In phase 1, we approximate the energy per particle using only the PNM energy:
\begin{equation}
    \left(\frac{E}{A}\right)^{(1)} (n,x) = \frac{E^{\text{bulk}}}{A}(n_n,0) \, .
\label{eq:energyphase1}
\end{equation}
In phase $2$, we consider additional finite-size contributions of the nuclear drop. 
With the Coulomb energy $E^{\text{C}}$ and the surface energy $E^{\text{S}}$, we write the energy per particle as
\begin{equation}
    \left(\frac{E}{A}\right)^{(2)} (n,x) = \frac{E^{\text{bulk}}}{A}(n,x) + \frac{E^{\text{C}}}{A}(n,x) + \frac{E^{\text{S}}}{A}(n,x) \, .
\label{eq:energyphase2}
\end{equation}
In both Eq.~\eqref{eq:energyphase1} as well as Eq.~\eqref{eq:energyphase2}, we describe the bulk part of the energy with the GP emulator based on the N$^3$LO results. Derivatives are calculated consistently from the GP.

The Coulomb contribution to the energy per particle is given by~\cite{Chamel_2008}
\begin{equation}
    \frac{E^{\text{C}}}{A}(n,x) = \frac{3}{5} \alpha_{\text{FS}} \left(\frac{4\pi}{3} \right)^{\frac{1}{3}}Z^{\frac{2}{3}}n^{\frac{1}{3}}x^{\frac{4}{3}} \left(1- \frac{3}{2}u^{\frac{1}{3}} + \frac{1}{2}u \right) \, ,
\label{eq:Coulombenergy}
\end{equation}
where $\alpha_{\text{FS}}$ is the fine structure constant and $u=V_0/V_{\text{W}}$ the volume fraction of the nucleus. The first term in Eq.~\eqref{eq:Coulombenergy} is the Coulomb energy of a nucleus with equally distributed protons. The contribution of the electrons to the Coulomb energy is encaptured by the latter two terms, the first of which is the lattice energy, assuming pointlike nuclei. The last term in Eq.~\eqref{eq:Coulombenergy} corrects for the finite size of the nuclei.

The surface contribution to the energy per particle can be expressed in terms of the surface tension $\sigma (x)$. With the surface area defined by the radius $R_0$ of the nuclear drop, we have~\cite{Tews_2017}
\begin{equation}
    \frac{E^{\text{S}}}{A}(n,x) = \frac{1}{A}\sigma (x) 4 \pi R_0^2 = \sigma (x) \left(\frac{36\pi x}{Z n^2} \right)^{\frac{1}{3}} \, .
\end{equation}
Following Ref.~\cite{Steiner_2008}, the isospin dependence of the surface tension can be modeled by expanding in the neutron excess $\delta$
\begin{equation}
    \sigma (x) = \sigma_0(1- \sigma_\delta(1-2x)^2 + \ldots) \, ,
\label{eq:surfacetension1}
\end{equation}
where the parameter $\sigma_\delta$ represents the surface symmetry energy. For the inner crust model, however, the surface tension has to vanish for $x\rightarrow 0$, since this implies the merge of phases $1$ and $2$. As proposed in Ref.~\cite{LLPR} one can thus modify the surface tension as
\begin{equation}
    \sigma (x) = \sigma_0 \frac{16+b}{x^{-3}+b+ (1-x)^{-3}} \, ,
\label{eq:surfacetension2}
\end{equation}
where $b=96/\sigma_\delta - 16$. For determining the surface tension, we follow the approach from Ref.~\cite{Tews_2017} and fit to experimental binding energies of nuclei. To this end, we employ Eq.~\eqref{eq:energyphase2} with the surface tension modeled by Eq.~\eqref{eq:surfacetension1}, and perform a least-squared optimization to the binding energies of the nuclei from the 2020 Atomic Mass Evaluation (AME 2020)~\cite{AME}. For nuclei, only the first term in Eq.~\eqref{eq:Coulombenergy} must be used, as the others stem from the existence of the electron gas. With this procedure, we obtain the following surface parameters:
\begin{equation}
    \sigma_0 = 0.84 \, \text{MeV\,fm}^{-2} \quad \text{and} \quad  \sigma_\delta = 2.27 \, \text{MeV\,fm}^{-2} \, .
\label{eq:surfaceparameters}
\end{equation}
This fit is best around $Z \approx 40$; very heavy and very light nuclei are less well approximated. Varying the surface parameters within a $10\%$ uncertainty reproduces nearly all binding energies for $15\leq Z \leq 80$. This domain coincides with the charge numbers relevant for the neutron star crust (see, e.g., \mbox{Refs. \cite{Pethick1995,Haensel:2007yy_NS1,Grams_2022_2}}). Also, Ref.~\cite{Tews_2017} studied the impact of variations in the surface tension to the inner crust and it was found that the uncertainty stemming from the surface parameters is sizable only for the chemical potentials at low densities. For the composition of the nuclear cluster, i.e., radii and mass numbers, variations in the surface tension only have a small effect compared to variations in the charge number (see below) and the uncertainty from the neutron matter EOS. Hence, we will continue with the surface parameters from Eq.~\eqref{eq:surfaceparameters} and neglect additional uncertainties from the surface tension. From the determined $\sigma_\delta$, we compute $b$ and employ the surface tension as defined in Eq.~\eqref{eq:surfacetension2} for the inner crust.

To calculate the EOS in $\beta$ equilibrium, the contributions of the electron gas have to be accounted for explicitly.
The energy density of a free relativistic electron gas $\mathcal{E}_{e}$ with density $n_{e}$ is given by~\cite{Chamel_2008}
\begin{equation}
    \mathcal{E}_{e}(n_{e}) = \frac{m_{e}^4}{8\pi^2}[x_r(2x_r^2+1)(x_r^2+1)^{1/2} - \ln(x_r+(x_r^2+1)^{1/2}) ] \, ,
\end{equation}
with $x_r = (3\pi^2n_{e}^{1/3})/m_{e}$ and $m_{e}=0.511 \mev$ is the electron mass. This energy density leads to a pressure given by 
\begin{equation}
    P_{e}(n_{e}) = n_{e}\frac{\partial \mathcal{E}_{e}}{\partial n_{e}} - \mathcal{E}_{e} \, ,
\end{equation}
and a chemical potential described by Eq.~\eqref{eq:electronchemicalpotential}.

\begin{figure}[t!]
    \centering
    \includegraphics[width=\linewidth]{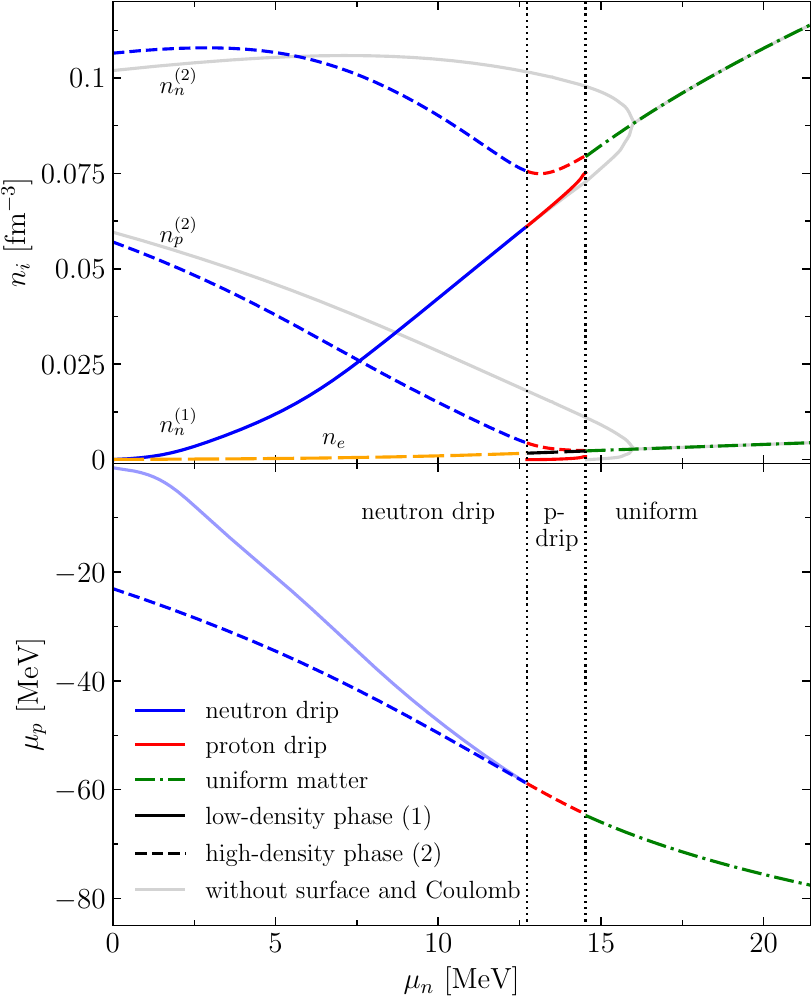}
    \caption{Phase coexistence in neutron star matter, obtained using the GP emulator based on the N$^3$LO results. Top panel: neutron and proton densities $n_n$ and $n_p$ as function of the neutron chemical potential $\mu_n$. The lower (higher) density phase is given by solid (dashed) lines. The colors indicate the different possibilities of coexistence. The electron density $n_e$ is shown in orange (neutron drip) and black (proton drip), respectively. The light gray lines show the results when surface and Coulomb corrections are not included~\cite{Keller_2024}.
    Bottom panel: corresponding proton chemical potential $\mu_p$. The unphysical line for the PNM phase is drawn in lighter blue.}
    \label{fig:phasecoexistence}
\end{figure}

Finally, before solving the equilibrium conditions, the charge number $Z$ of the nucleus has to be determined. In theory, this can be done by minimizing the energy within the Wigner-Seitz cell. A rigorous treatment with the inclusion of shell effects is, however, involved. Hence, we follow Ref.~\cite{Tews_2017} and assume a closed proton shell with $Z=40$ for all densities. In Ref.~\cite{Tews_2017}, the author also varied the proton number in a range $Z=28\!-\!50$, which includes the two neighboring shell closures and covers the results of recent studies (see, e.g., Ref.~\cite{Grams_2022_2}). It is only near the crust-core boundary, where the existence of pasta phases can lead to significantly larger charge numbers. While variations in $Z$ of course define the size and mass number of the Wigner-Seitz cell, it was found that they do not significantly impact the chemical potentials and thus the EOS. Since we are primarily interested in the EOS and not in determining the exact composition of the Wigner-Seitz cell, we continue with this choice of a constant~$Z$.

With these definitions, we model the neutron star crust as a Coulomb lattice of Wigner-Seitz cells. We solve the equilibrium conditions for a single cell on a grid of neutron chemical potentials $\mu_n$. Neutron drip, and thus the inner crust, starts when the neutron chemical potential reaches $\mu_n \geq 0$. In our model, we also allow for the possibility of proton drip, which sets in once Eq.~\eqref{eq:neutrondrip} is not satisfied anymore. It is then energetically favorable for protons to populate the previously PNM phase, such that $n_p^{(2)}>0$. For equilibrium, Eq.~\eqref{eq:neutrondrip} thus becomes $\mu_p^{(1)} = \mu_p^{(2)}$ and we have to adapt the energy per particle in this phase, Eq.~\eqref{eq:energyphase1}, to
\begin{equation}
    \left(\frac{E}{A}\right)^{(1)} (n,x) = \frac{E^{\text{bulk}}}{A}(n,x) \, .
\end{equation}
Here, we do not account for additional Coulomb and surface effects, as they are negligible compared to the bulk contribution. Pasta phases, which may be formed near the crust-core boundary, cannot be considered in our model. Their inclusion would require a consistent computation of proton numbers as well as a more rigorous treatment of lattice effects and the surface energy, allowing for non-spherical Wigner-Seitz cells.

The solutions to the inner crust equilibrium conditions are shown in Fig.~\ref{fig:phasecoexistence}. The upper panel shows the neutron and proton densities of both phases as a function of neutron chemical potential. The light gray lines show the results when surface and Coulomb corrections are not included~\cite{Keller_2024}. Comparing these results, the inclusion of finite-size terms leads to a more rapid decrease of the neutron and proton densities in the high-density phase. Since Coulomb and surface effects however increase the proton chemical potential, we still find a region where proton drip is possible. At $\mu_n \approx \SI{12.7}{MeV}$ the proton chemical potentials of both phases match (lower panel) and protons start to drip out of the high-density phase. The two nuclear phases coexist until $\mu_n \approx \SI{14.5}{MeV}$. Here, the electron density matches the proton density of the high-density phase and uniform matter begins; otherwise $\beta$ equilibrium cannot be satisfied anymore. This transition between the neutron star inner crust and uniform matter in the core is a first-order phase transition with a small density jump in the lower-density phase.

With the densities in both phases determined, the total baryon density $n$ in the Wigner-Seitz cell can be obtained via
\begin{equation}
    n = (1-u)(n_n^{(1)}+n_p^{(1)}) + u(n_n^{(2)}+n_p^{(2)}) \, .
\label{eq:baryondensity}
\end{equation}
The volume fraction $u$ follows directly from enforcing charge neutrality, $Q=V_0n_p^{(2)}-V_{\text{W}}n_{e} + (V_{\text{W}}-V_0)n_p^{(1)}=0$. To obtain the EOS, $P = P(n)$, we compute the total pressure as \mbox{$P=P^{(1)}+P_{e}=P^{(2)}+P_{e}$}.

\subsection{Crust EOS including EFT uncertainties}

\begin{figure}[t!]
    \centering
    \includegraphics[width=\linewidth]{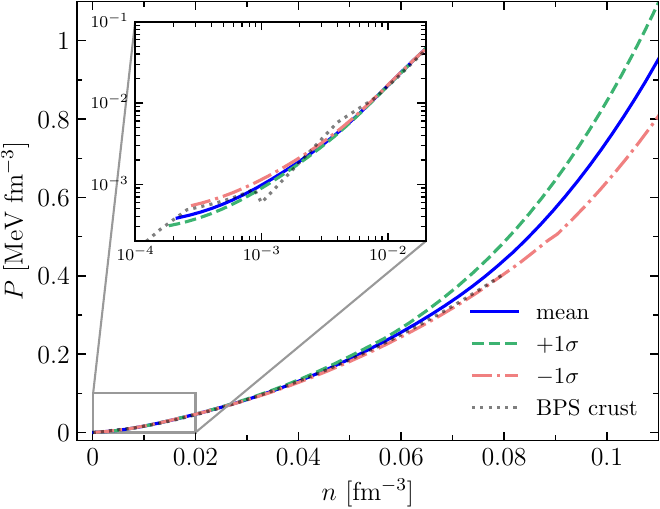}
    \caption{Pressure of neutron star matter as function of baryon density from the crust to the outer core. The blue line corresponds to the GP mean for the N$^3$LO results and the $\pm\sigma$ to matching the upper and lower 68\% credibility range of the uniform pressure, see Eq.~\eqref{eq:crustsubstitutions}. For comparison, we also show the BPS crust EOS~\cite{BPS}.}
    \label{fig:EOS}
\end{figure}

For the construction of uncertainty bands of the neutron star crust EOS, we focus on the EFT uncertainties estimated with the 2D GP developed in Sec.~\ref{sec:2DGP}. With this, we can compute the standard deviation to the nuclear energy per particle or any derivative at arbitrary density and proton fraction. While the GP also allows drawing individual samples, constructing a crust EOS from such samples is not always possible as they may exhibit unphysical variations. Thus, we restrict our analysis to the standard deviations, which are well-behaved but still contain the most important statistical information. In addition to the mean line considered in Sec.~\ref{sec:crusttheory}, we consider two representative cases: one treating the upper boundary of the 68\% credibility band ($+\sigma$) and one treating the lower boundary ($-\sigma$).

In particular, we make the substitution 
\begin{equation}
    \frac{E^{\text{bulk}}}{A}(n,x) \rightarrow \frac{E^{\text{bulk}}}{A}(n,x) \pm \sigma \bigg(\frac{E^{\text{bulk}}}{A}\bigg)(n,x) \, ,
\label{eq:crustsubstitutions}
\end{equation}
where $\sigma$ denotes the standard deviation extracted from the GP. Using these energy functionals, we newly fit the surface parameters as described before. For the construction of the crust, we make the analogous substitutions for the pressure and the nucleon chemical potentials:
\begin{align}
   P(n,x) &\rightarrow P(n,x) \pm \sigma (P)(n,x) \, , \\ 
   \mu_\tau(n,x) &\rightarrow  \mu_\tau(n,x) \pm \sigma  (\mu_\tau)(n,x) \, .
\end{align}
With these new expressions, the equilibrium conditions are again solved as described in Sec.~\ref{sec:crusttheory}.

\begin{figure}[t!]
    \centering
    \includegraphics[width=\linewidth]{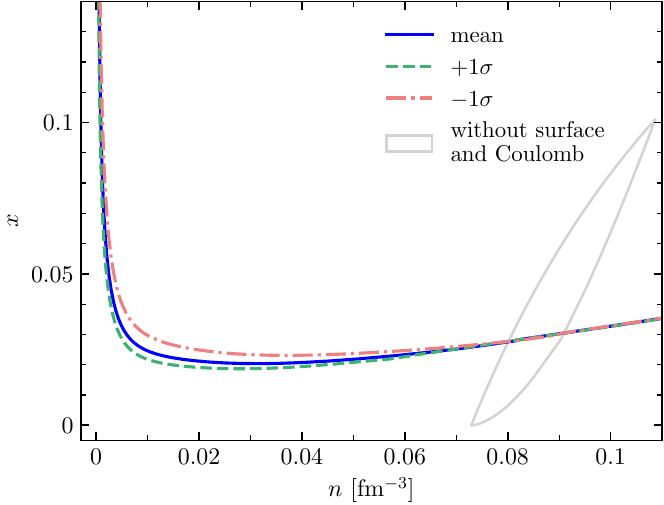}
    \caption{Total proton fraction within the Wigner-Seitz cell as a function of baryon density. The blue line corresponds to the GP mean for the N$^3$LO results and the $\pm\sigma$ to matching the upper and lower 68\% credibility range of the uniform pressure, see Eq.~\eqref{eq:crustsubstitutions}. The gray outlined region marks the proton drip phase when surface and Coulomb corrections are not included~\cite{Keller_2024}.}
    \label{fig:neutronstarphasediagram}
\end{figure}

We present the inner crust EOSs consistent with the N$^3$LO results and their EFT uncertainties in Fig.~\ref{fig:EOS}. As discussed above, we expect the effects on the EOS from pasta phases or from improvements over the Wigner-Seitz cell approximation made here to be small compared to the EFT uncertainties. The uncertainty band spanned by the mean and $\pm\sigma$ boundaries match smoothly to the BPS crust EOS and this matching continues to low densities (see inset). In addition, for $n \lesssim 0.016 \fmiq$, we find an inversion of the upper and lower boundary from the uniform band. In Fig.~\ref{fig:neutronstarphasediagram}, we show the total proton fraction as a function of baryon density for the three considered cases. We observe that the relatively stiffer (softer) EOS leads to smaller (larger) proton fractions in the inner crust. Notably, the differences are negligible in the uniform-matter phase and all three energy functionals yield nearly identical results.

The gray outlined region in Fig.~\ref{fig:neutronstarphasediagram} marks the proton drip phase when surface and Coulomb corrections are not included~\cite{Keller_2024}. The inclusion of finite-size terms reduces the range of densities where proton drip can occur, yet for all three constructed crust EOSs, we find proton drip before the system reaches the uniform phase. For the soft ($-\sigma$), mean, and stiff ($+\sigma$) EOS, proton drip is possible for densities \mbox{$(0.077\!\mathbin{-}\!0.088)\fmiq$}, \mbox{$(0.068\!\mathbin{-}\!0.082)\fmiq$}, and \mbox{$(0.058\!\mathbin{-}\!0.062)\fmiq$}, respectively. In particular, for the upper boundary of the $68\%$ credibility range, the proton drip region almost disappears; incorporating additional effects from nuclear clusters or pasta phases near the crust-core boundary, may then result in the proton drip phase being jumped over. The obtained crust-core transition densities between $0.062$ and $0.088 \fmiq$ agree well with other studies (see, e.g., Refs.~\cite{Hebeler_2013,Tews_2017,Grams_2022_2}), although $0.062 \fmiq$ is more on the lower side of the range of commonly obtained values.

\section{Summary and outlook}
\label{sec:summary}

In this work, we combined recent state-of-the art chiral EFT calculations of asymmetric nuclear matter with a Bayesian framework for modeling the EFT truncation uncertainties, to obtain the neutron star EOS up to $2n_0$ with EFT uncertainties. Recently, in Ref.~\cite{Melendez:2019izc_EFT} a Bayesian truncation error model was proposed, formalizing the expected EFT convergence behavior and correlations along independent variables using Gaussian processes. In Refs.~\cite{Drischler:2020hwi_EOS,Drischler:2020yad_matter}, this framework was applied to the PNM and SNM EOS separately. In this paper, we extended the quantification of EFT truncation uncertainties to asymmetric nuclear matter. To this end, we constructed a 2D GP, with a kernel that accounts for correlations both along the density $n$ and the proton fraction $x$.

To justify the application of a smooth and stationary kernel, we investigated the order-by-order corrections to the energy per particle in detail. We found it beneficial to include the effects of 3N contributions on the expansion coefficients entering at N$^2$LO through an improved $x$-dependent reference energy. This restores a common scale of the coefficients across all proton fractions and at high densities. In addition, we thus studied the sensitivity of the hyperparameters to the reference energy, the training set, and the use of nonuniform priors. This led to leaving out higher proton fractions, which are symmetric around $x = 0.5$, to not add bias on the correlation length, and to employ a conjugate inverse-$\chi^2$ prior for the scale of the coefficients, incorporating the physical bias of a well-behaved EFT convergence. We validated the description of the order-by-order corrections by the 2D GP with model-checking diagnostics.

Using this setup, we provided EFT uncertainties for asymmetric nuclear matter that account for correlations in both density and proton fraction. With the 2D GP, we calculated the neutron star EOS in $\beta$ equilibrium with associated EFT uncertainties without the need for phenomenological interpolations between PNM and SNM. In addition to the energy per particle, the pressure, and the proton fraction, we also presented results for the nucleon chemical potentials. At N$^3$LO, the proton fraction in $\beta$ equilibrium increases monotonously with density and reaches up to $7.5\%$ at $2 n_0$.

We further extended the EOS to the neutron star crust regime of non-uniform matter. For modeling the inner crust, we employed a compressible liquid drop model with the Wigner-Seitz approximation. Taking the limits of the $68\%$ credibility intervals from the 2D GP, we solved the crust equilibrium conditions for three cases, spanning a representative area for the neutron star EOS to account for EFT truncation uncertainties in the low-density regime. Our calculations show that proton drip is possible in all three cases, although the range of densities where high-density phases can coexist varies. We found the crust-core transition densities between \mbox{$(0.062\!-\! 0.088) \fmiq$}. Near this crust-core boundary, the appearance of nonspherical nuclear shapes is expected to have a significant impact on the cluster composition; this cannot be studied in our current model and remains for future work.  However, for macroscopic thermodynamic properties like the EOS, previous studies have shown that the compressible liquid drop model is in good agreement with more sophisticated approaches to the finite-size modeling~\cite{Grams:2022ojr}.

A great advantage of our neutron star crust and outer core EOS lies in the possibility to extend the 2D GP for the EOS and for EFT uncertainties to finite temperatures (see Ref.~\cite{Keller:2022crb}). In addition, the EFT truncation uncertainties discussed in this work can be combined with the (smaller) many-body uncertainties arising from the many-body perturbation theory expansion~\cite{Svensson:2025jde}. Finally, we will incorporate our consistent crust EOSs in future work in the astrophysics Bayesian inference framework using these chiral EFT results in the uniform phase~\cite{Rutherford:2024srk}.

\begin{acknowledgments}
We thank Jonas Keller, Melissa Mendes, Chris Pethick, and Isak Svensson for useful discussions. This work was supported in part by the European Research Council (ERC) under the European Union’s Horizon 2020 research and innovation programme (Grant Agreement No.~101020842) and by the LOEWE Top Professorship LOEWE/4a/519/05.00.002(0014)98 by the State of Hesse.
\end{acknowledgments}

\section*{Data availability}

The data that support the findings of this article are openly
available~\cite{zenodo}.

\appendix
\section*{Appendix: Expansion coefficients\\ for $\erefB$}
\addcontentsline{toc}{section}{Appendix: Expansion coefficients\\ for $\erefB$}

\begin{figure*}[t!]
    \centering
    \includegraphics[width=0.8\linewidth]{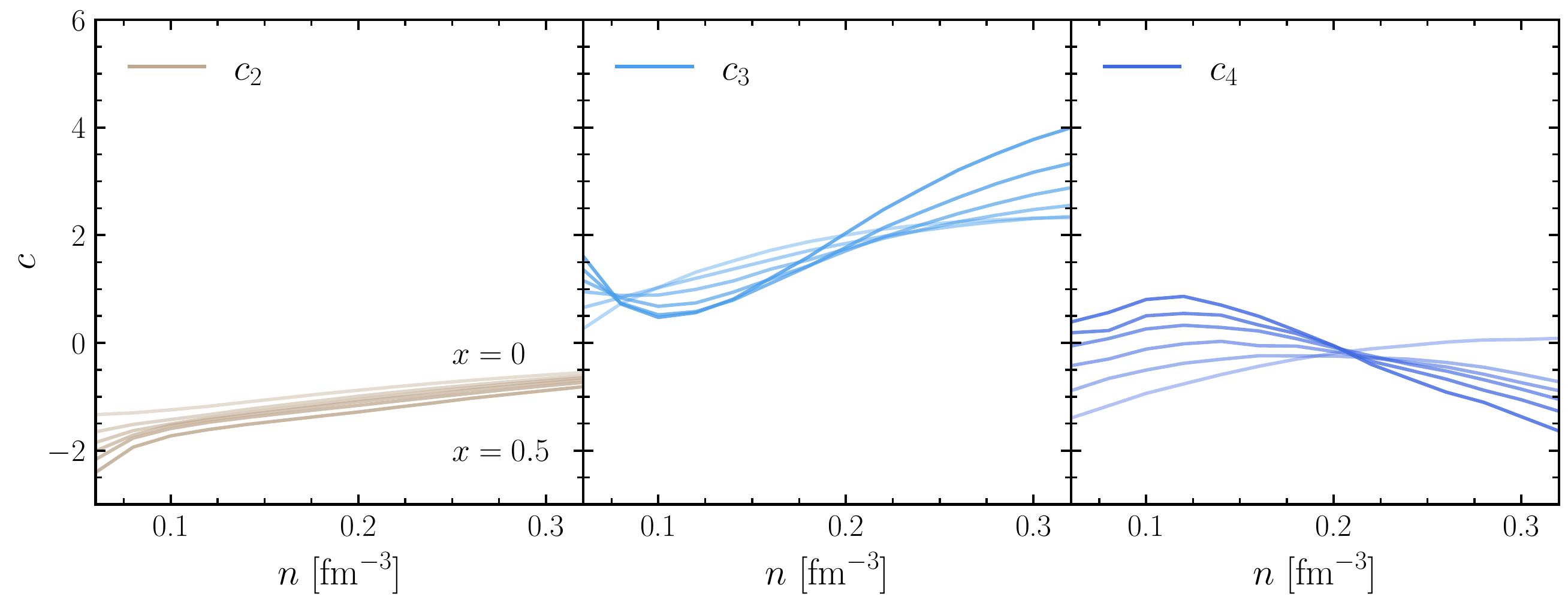}
    \caption{Expansion coefficients at NLO ($c_2$, left), N$^2$LO ($c_3$, middle), and N$^3$LO ($c_4$, right panel) extracted from the asymmetric matter calculations with reference energy $\erefB$ as a function of density $n$. Results are shown for different proton fractions from $x=0$ to $x=0.5$ in steps of $0.1$ (from light to darker).}
    \label{fig:c_n_final}
\end{figure*}

\begin{figure*}[t!]
    \centering
    \includegraphics[width=0.8\linewidth]{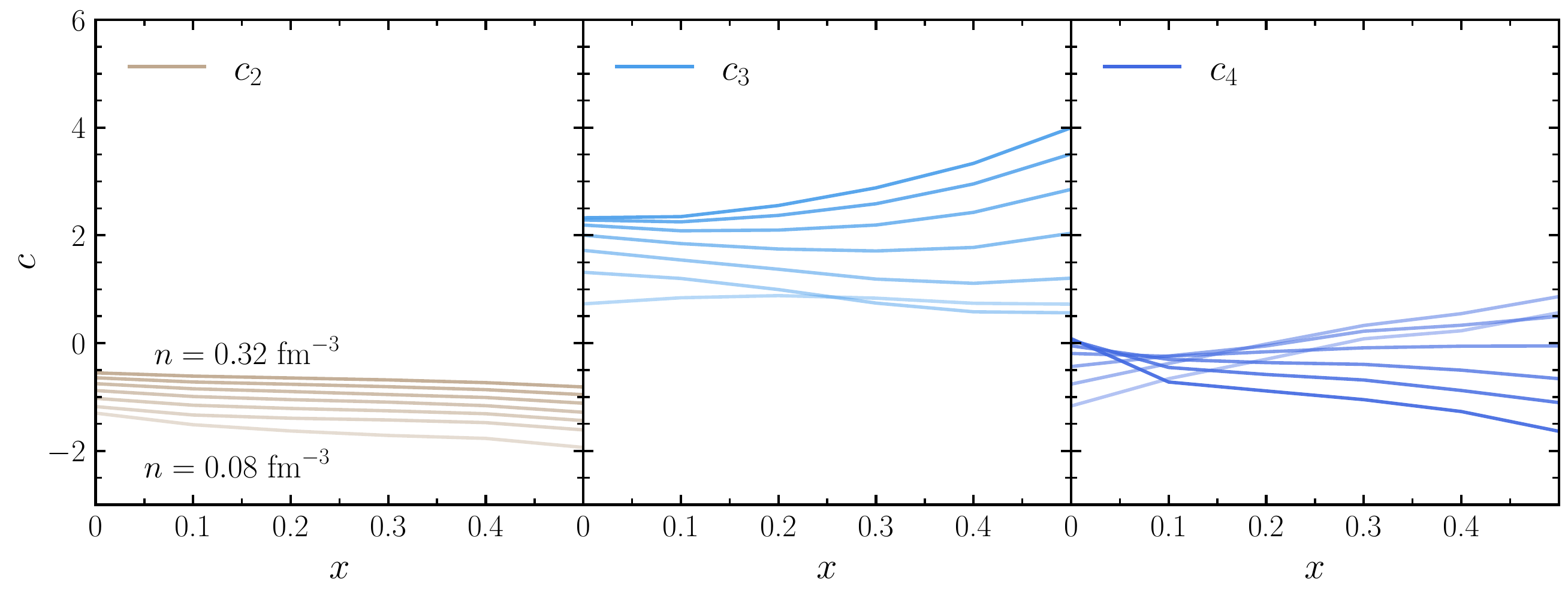}
    \caption{Expansion coefficients analogous to Fig.~\ref{fig:c_n_final}, here as function of proton fraction $x$ shown for selected fixed densities between $n = 0.08 \fmiq$ and $n = 0.32 \fmiq$ in steps of $0.04 \fmiq$ (from light to darker).}
    \label{fig:c_x_final}
\end{figure*}

In Figs.~\ref{fig:c_n_final} and~\ref{fig:c_x_final} we show the expansion coefficients $c_2(n,x)$, $c_3(n,x)$, and $c_4(n,x)$ using the proton-fraction-dependent reference energy $\erefB$ introduced in Eq.~\eqref{eq:newrefenergy}, which is used for our main results. With this, the impact of 3N contributions at N$^2$LO and higher order is better accounted for compared to using the reference energy $\erefBQ$. As shown in the figure, the expansion coefficients are in the range $c_i(n, x) \approx [-2, 4]$, and by only considering proton fractions up to $x = 0.5$ we can ensure stationarity along both variables $n$ and $x$.

\bibliography{literature}

\begin{thebibliography}{49}%
\makeatletter
\providecommand \@ifxundefined [1]{%
 \@ifx{#1\undefined}
}%
\providecommand \@ifnum [1]{%
 \ifnum #1\expandafter \@firstoftwo
 \else \expandafter \@secondoftwo
 \fi
}%
\providecommand \@ifx [1]{%
 \ifx #1\expandafter \@firstoftwo
 \else \expandafter \@secondoftwo
 \fi
}%
\providecommand \natexlab [1]{#1}%
\providecommand \enquote  [1]{``#1''}%
\providecommand \bibnamefont  [1]{#1}%
\providecommand \bibfnamefont [1]{#1}%
\providecommand \citenamefont [1]{#1}%
\providecommand \href@noop [0]{\@secondoftwo}%
\providecommand \href [0]{\begingroup \@sanitize@url \@href}%
\providecommand \@href[1]{\@@startlink{#1}\@@href}%
\providecommand \@@href[1]{\endgroup#1\@@endlink}%
\providecommand \@sanitize@url [0]{\catcode `\\12\catcode `\$12\catcode `\&12\catcode `\#12\catcode `\^12\catcode `\_12\catcode `\%12\relax}%
\providecommand \@@startlink[1]{}%
\providecommand \@@endlink[0]{}%
\providecommand \url  [0]{\begingroup\@sanitize@url \@url }%
\providecommand \@url [1]{\endgroup\@href {#1}{\urlprefix }}%
\providecommand \urlprefix  [0]{URL }%
\providecommand \Eprint [0]{\href }%
\providecommand \doibase [0]{https://doi.org/}%
\providecommand \selectlanguage [0]{\@gobble}%
\providecommand \bibinfo  [0]{\@secondoftwo}%
\providecommand \bibfield  [0]{\@secondoftwo}%
\providecommand \translation [1]{[#1]}%
\providecommand \BibitemOpen [0]{}%
\providecommand \bibitemStop [0]{}%
\providecommand \bibitemNoStop [0]{.\EOS\space}%
\providecommand \EOS [0]{\spacefactor3000\relax}%
\providecommand \BibitemShut  [1]{\csname bibitem#1\endcsname}%
\let\auto@bib@innerbib\@empty
\bibitem [{\citenamefont {Hebeler}\ \emph {et~al.}(2013)\citenamefont {Hebeler}, \citenamefont {Lattimer}, \citenamefont {Pethick},\ and\ \citenamefont {Schwenk}}]{Hebeler_2013}%
  \BibitemOpen
  \bibfield  {author} {\bibinfo {author} {\bibfnamefont {K.}~\bibnamefont {Hebeler}}, \bibinfo {author} {\bibfnamefont {J.~M.}\ \bibnamefont {Lattimer}}, \bibinfo {author} {\bibfnamefont {C.~J.}\ \bibnamefont {Pethick}},\ and\ \bibinfo {author} {\bibfnamefont {A.}~\bibnamefont {Schwenk}},\ }\bibfield  {title} {\bibinfo {title} {{Equation of state and neutron star properties constrained by nuclear physics and observation}},\ }\href {https://doi.org/10.1088/0004-637X/773/1/11} {\bibfield  {journal} {\bibinfo  {journal} {Astrophys. J.}\ }\textbf {\bibinfo {volume} {773}},\ \bibinfo {pages} {11} (\bibinfo {year} {2013})}\BibitemShut {NoStop}%
\bibitem [{\citenamefont {Yasin}\ \emph {et~al.}(2020)\citenamefont {Yasin}, \citenamefont {Sch\"afer}, \citenamefont {Arcones},\ and\ \citenamefont {Schwenk}}]{Yasin:2018ckc}%
  \BibitemOpen
  \bibfield  {author} {\bibinfo {author} {\bibfnamefont {H.}~\bibnamefont {Yasin}}, \bibinfo {author} {\bibfnamefont {S.}~\bibnamefont {Sch\"afer}}, \bibinfo {author} {\bibfnamefont {A.}~\bibnamefont {Arcones}},\ and\ \bibinfo {author} {\bibfnamefont {A.}~\bibnamefont {Schwenk}},\ }\bibfield  {title} {\bibinfo {title} {{Equation of state effects in core-collapse supernovae}},\ }\href {https://doi.org/10.1103/PhysRevLett.124.092701} {\bibfield  {journal} {\bibinfo  {journal} {Phys. Rev. Lett.}\ }\textbf {\bibinfo {volume} {124}},\ \bibinfo {pages} {092701} (\bibinfo {year} {2020})}\BibitemShut {NoStop}%
\bibitem [{\citenamefont {Schneider}\ \emph {et~al.}(2019)\citenamefont {Schneider}, \citenamefont {Roberts}, \citenamefont {Ott},\ and\ \citenamefont {O'Connor}}]{Schneider:2019shi}%
  \BibitemOpen
  \bibfield  {author} {\bibinfo {author} {\bibfnamefont {A.~S.}\ \bibnamefont {Schneider}}, \bibinfo {author} {\bibfnamefont {L.~F.}\ \bibnamefont {Roberts}}, \bibinfo {author} {\bibfnamefont {C.~D.}\ \bibnamefont {Ott}},\ and\ \bibinfo {author} {\bibfnamefont {E.}~\bibnamefont {O'Connor}},\ }\bibfield  {title} {\bibinfo {title} {{Equation of state effects in the core collapse of a $20$-$M_\odot$ star}},\ }\href {https://doi.org/10.1103/PhysRevC.100.055802} {\bibfield  {journal} {\bibinfo  {journal} {Phys. Rev. C}\ }\textbf {\bibinfo {volume} {100}},\ \bibinfo {pages} {055802} (\bibinfo {year} {2019})}\BibitemShut {NoStop}%
\bibitem [{\citenamefont {Lattimer}(2021)}]{Lattimer2021}%
  \BibitemOpen
  \bibfield  {author} {\bibinfo {author} {\bibfnamefont {J.~M.}\ \bibnamefont {Lattimer}},\ }\bibfield  {title} {\bibinfo {title} {{Neutron Stars and the Nuclear Matter Equation of State}},\ }\href {https://doi.org/10.1146/annurev-nucl-102419-124827} {\bibfield  {journal} {\bibinfo  {journal} {Annu. Rev. Nucl. Part. Sci.}\ }\textbf {\bibinfo {volume} {71}},\ \bibinfo {pages} {433} (\bibinfo {year} {2021})}\BibitemShut {NoStop}%
\bibitem [{\citenamefont {Jacobi}\ \emph {et~al.}(2023)\citenamefont {Jacobi}, \citenamefont {Guercilena}, \citenamefont {Huth}, \citenamefont {Ricigliano}, \citenamefont {Arcones},\ and\ \citenamefont {Schwenk}}]{Jacobi:2023olu}%
  \BibitemOpen
  \bibfield  {author} {\bibinfo {author} {\bibfnamefont {M.}~\bibnamefont {Jacobi}}, \bibinfo {author} {\bibfnamefont {F.~M.}\ \bibnamefont {Guercilena}}, \bibinfo {author} {\bibfnamefont {S.}~\bibnamefont {Huth}}, \bibinfo {author} {\bibfnamefont {G.}~\bibnamefont {Ricigliano}}, \bibinfo {author} {\bibfnamefont {A.}~\bibnamefont {Arcones}},\ and\ \bibinfo {author} {\bibfnamefont {A.}~\bibnamefont {Schwenk}},\ }\bibfield  {title} {\bibinfo {title} {{Effects of nuclear matter properties in neutron star mergers}},\ }\href {https://doi.org/10.1093/mnras/stad3738} {\bibfield  {journal} {\bibinfo  {journal} {Mon. Not. Roy. Astron. Soc.}\ }\textbf {\bibinfo {volume} {527}},\ \bibinfo {pages} {8812} (\bibinfo {year} {2023})}\BibitemShut {NoStop}%
\bibitem [{\citenamefont {Chatziioannou}\ \emph {et~al.}(2025)\citenamefont {Chatziioannou}, \citenamefont {Cromartie}, \citenamefont {Gandolfi}, \citenamefont {Tews}, \citenamefont {Radice}, \citenamefont {Steiner},\ and\ \citenamefont {Watts}}]{EOSreview}%
  \BibitemOpen
  \bibfield  {author} {\bibinfo {author} {\bibfnamefont {K.}~\bibnamefont {Chatziioannou}}, \bibinfo {author} {\bibfnamefont {H.~T.}\ \bibnamefont {Cromartie}}, \bibinfo {author} {\bibfnamefont {S.}~\bibnamefont {Gandolfi}}, \bibinfo {author} {\bibfnamefont {I.}~\bibnamefont {Tews}}, \bibinfo {author} {\bibfnamefont {D.}~\bibnamefont {Radice}}, \bibinfo {author} {\bibfnamefont {A.~W.}\ \bibnamefont {Steiner}},\ and\ \bibinfo {author} {\bibfnamefont {A.~L.}\ \bibnamefont {Watts}},\ }\bibfield  {title} {\bibinfo {title} {Neutron stars and the dense matter equation of state},\ }\href {https://doi.org/10.1103/ymsq-cfcw} {\bibfield  {journal} {\bibinfo  {journal} {Rev. Mod. Phys.}\ }\textbf {\bibinfo {volume} {97}},\ \bibinfo {pages} {045007} (\bibinfo {year} {2025})}\BibitemShut {NoStop}%
\bibitem [{\citenamefont {Epelbaum}\ \emph {et~al.}(2009)\citenamefont {Epelbaum}, \citenamefont {Hammer},\ and\ \citenamefont {Mei\ss{}ner}}]{Epelbaum_et_al_2009}%
  \BibitemOpen
  \bibfield  {author} {\bibinfo {author} {\bibfnamefont {E.}~\bibnamefont {Epelbaum}}, \bibinfo {author} {\bibfnamefont {H.-W.}\ \bibnamefont {Hammer}},\ and\ \bibinfo {author} {\bibfnamefont {U.-G.}\ \bibnamefont {Mei\ss{}ner}},\ }\bibfield  {title} {\bibinfo {title} {Modern theory of nuclear forces},\ }\href {https://doi.org/10.1103/RevModPhys.81.1773} {\bibfield  {journal} {\bibinfo  {journal} {Rev. Mod. Phys.}\ }\textbf {\bibinfo {volume} {81}},\ \bibinfo {pages} {1773} (\bibinfo {year} {2009})}\BibitemShut {NoStop}%
\bibitem [{\citenamefont {Machleidt}\ and\ \citenamefont {Entem}(2011)}]{Machleidt_Entem_2011}%
  \BibitemOpen
  \bibfield  {author} {\bibinfo {author} {\bibfnamefont {R.}~\bibnamefont {Machleidt}}\ and\ \bibinfo {author} {\bibfnamefont {D.~R.}\ \bibnamefont {Entem}},\ }\bibfield  {title} {\bibinfo {title} {Chiral effective field theory and nuclear forces},\ }\href {https://doi.org/https://doi.org/10.1016/j.physrep.2011.02.001} {\bibfield  {journal} {\bibinfo  {journal} {Phys. Rep.}\ }\textbf {\bibinfo {volume} {503}},\ \bibinfo {pages} {1} (\bibinfo {year} {2011})}\BibitemShut {NoStop}%
\bibitem [{\citenamefont {Hebeler}\ and\ \citenamefont {Schwenk}(2010)}]{HebelerSchwenk2010}%
  \BibitemOpen
  \bibfield  {author} {\bibinfo {author} {\bibfnamefont {K.}~\bibnamefont {Hebeler}}\ and\ \bibinfo {author} {\bibfnamefont {A.}~\bibnamefont {Schwenk}},\ }\bibfield  {title} {\bibinfo {title} {Chiral three-nucleon forces and neutron matter},\ }\href {https://doi.org/10.1103/PhysRevC.82.014314} {\bibfield  {journal} {\bibinfo  {journal} {Phys. Rev. C}\ }\textbf {\bibinfo {volume} {82}},\ \bibinfo {pages} {014314} (\bibinfo {year} {2010})}\BibitemShut {NoStop}%
\bibitem [{\citenamefont {Tews}\ \emph {et~al.}(2013)\citenamefont {Tews}, \citenamefont {Kr{\"u}ger}, \citenamefont {Hebeler},\ and\ \citenamefont {Schwenk}}]{Tews13N3LO}%
  \BibitemOpen
  \bibfield  {author} {\bibinfo {author} {\bibfnamefont {I.}~\bibnamefont {Tews}}, \bibinfo {author} {\bibfnamefont {T.}~\bibnamefont {Kr{\"u}ger}}, \bibinfo {author} {\bibfnamefont {K.}~\bibnamefont {Hebeler}},\ and\ \bibinfo {author} {\bibfnamefont {A.}~\bibnamefont {Schwenk}},\ }\bibfield  {title} {\bibinfo {title} {{Neutron Matter at Next-to-Next-to-Next-to-Leading Order in Chiral Effective Field Theory}},\ }\href {https://doi.org/10.1103/PhysRevLett.110.032504} {\bibfield  {journal} {\bibinfo  {journal} {Phys. Rev. Lett.}\ }\textbf {\bibinfo {volume} {110}},\ \bibinfo {pages} {032504} (\bibinfo {year} {2013})}\BibitemShut {NoStop}%
\bibitem [{\citenamefont {Holt}\ \emph {et~al.}(2013)\citenamefont {Holt}, \citenamefont {Kaiser},\ and\ \citenamefont {Weise}}]{Holt13PPNP}%
  \BibitemOpen
  \bibfield  {author} {\bibinfo {author} {\bibfnamefont {J.~W.}\ \bibnamefont {Holt}}, \bibinfo {author} {\bibfnamefont {N.}~\bibnamefont {Kaiser}},\ and\ \bibinfo {author} {\bibfnamefont {W.}~\bibnamefont {Weise}},\ }\bibfield  {title} {\bibinfo {title} {{Nuclear chiral dynamics and thermodynamics}},\ }\href {https://doi.org/10.1016/j.ppnp.2013.08.001} {\bibfield  {journal} {\bibinfo  {journal} {Prog. Part. Nucl. Phys.}\ }\textbf {\bibinfo {volume} {73}},\ \bibinfo {pages} {35} (\bibinfo {year} {2013})}\BibitemShut {NoStop}%
\bibitem [{\citenamefont {Carbone}\ \emph {et~al.}(2013)\citenamefont {Carbone}, \citenamefont {Polls},\ and\ \citenamefont {Rios}}]{Carb13nm}%
  \BibitemOpen
  \bibfield  {author} {\bibinfo {author} {\bibfnamefont {A.}~\bibnamefont {Carbone}}, \bibinfo {author} {\bibfnamefont {A.}~\bibnamefont {Polls}},\ and\ \bibinfo {author} {\bibfnamefont {A.}~\bibnamefont {Rios}},\ }\bibfield  {title} {\bibinfo {title} {{Symmetric nuclear matter with chiral three-nucleon forces in the self-consistent Green's functions approach}},\ }\href {https://doi.org/10.1103/PhysRevC.88.044302} {\bibfield  {journal} {\bibinfo  {journal} {Phys. Rev. C}\ }\textbf {\bibinfo {volume} {88}},\ \bibinfo {pages} {044302} (\bibinfo {year} {2013})}\BibitemShut {NoStop}%
\bibitem [{\citenamefont {Hagen}\ \emph {et~al.}(2014)\citenamefont {Hagen}, \citenamefont {Papenbrock}, \citenamefont {Ekstr{\"o}m}, \citenamefont {Wendt}, \citenamefont {Baardsen}, \citenamefont {Gandolfi}, \citenamefont {Hjorth-Jensen},\ and\ \citenamefont {Horowitz}}]{Hage14ccnm}%
  \BibitemOpen
  \bibfield  {author} {\bibinfo {author} {\bibfnamefont {G.}~\bibnamefont {Hagen}}, \bibinfo {author} {\bibfnamefont {T.}~\bibnamefont {Papenbrock}}, \bibinfo {author} {\bibfnamefont {A.}~\bibnamefont {Ekstr{\"o}m}}, \bibinfo {author} {\bibfnamefont {K.~A.}\ \bibnamefont {Wendt}}, \bibinfo {author} {\bibfnamefont {G.}~\bibnamefont {Baardsen}}, \bibinfo {author} {\bibfnamefont {S.}~\bibnamefont {Gandolfi}}, \bibinfo {author} {\bibfnamefont {M.}~\bibnamefont {Hjorth-Jensen}},\ and\ \bibinfo {author} {\bibfnamefont {C.~J.}\ \bibnamefont {Horowitz}},\ }\bibfield  {title} {\bibinfo {title} {{Coupled-cluster calculations of nucleonic matter}},\ }\href {https://doi.org/10.1103/PhysRevC.89.014319} {\bibfield  {journal} {\bibinfo  {journal} {Phys. Rev. C}\ }\textbf {\bibinfo {volume} {89}},\ \bibinfo {pages} {014319} (\bibinfo {year} {2014})}\BibitemShut {NoStop}%
\bibitem [{\citenamefont {Coraggio}\ \emph {et~al.}(2014)\citenamefont {Coraggio}, \citenamefont {Holt}, \citenamefont {Itaco}, \citenamefont {Machleidt}, \citenamefont {Marcucci},\ and\ \citenamefont {Sammarruca}}]{Coraggio_et_al_2014}%
  \BibitemOpen
  \bibfield  {author} {\bibinfo {author} {\bibfnamefont {L.}~\bibnamefont {Coraggio}}, \bibinfo {author} {\bibfnamefont {J.~W.}\ \bibnamefont {Holt}}, \bibinfo {author} {\bibfnamefont {N.}~\bibnamefont {Itaco}}, \bibinfo {author} {\bibfnamefont {R.}~\bibnamefont {Machleidt}}, \bibinfo {author} {\bibfnamefont {L.~E.}\ \bibnamefont {Marcucci}},\ and\ \bibinfo {author} {\bibfnamefont {F.}~\bibnamefont {Sammarruca}},\ }\bibfield  {title} {\bibinfo {title} {Nuclear-matter equation of state with consistent two- and three-body perturbative chiral interactions},\ }\href {https://doi.org/10.1103/PhysRevC.89.044321} {\bibfield  {journal} {\bibinfo  {journal} {Phys. Rev. C}\ }\textbf {\bibinfo {volume} {89}},\ \bibinfo {pages} {044321} (\bibinfo {year} {2014})}\BibitemShut {NoStop}%
\bibitem [{\citenamefont {Wellenhofer}\ \emph {et~al.}(2014)\citenamefont {Wellenhofer}, \citenamefont {Holt}, \citenamefont {Kaiser},\ and\ \citenamefont {Weise}}]{PhysRevC.89.064009}%
  \BibitemOpen
  \bibfield  {author} {\bibinfo {author} {\bibfnamefont {C.}~\bibnamefont {Wellenhofer}}, \bibinfo {author} {\bibfnamefont {J.~W.}\ \bibnamefont {Holt}}, \bibinfo {author} {\bibfnamefont {N.}~\bibnamefont {Kaiser}},\ and\ \bibinfo {author} {\bibfnamefont {W.}~\bibnamefont {Weise}},\ }\bibfield  {title} {\bibinfo {title} {Nuclear thermodynamics from chiral low-momentum interactions},\ }\href {https://doi.org/10.1103/PhysRevC.89.064009} {\bibfield  {journal} {\bibinfo  {journal} {Phys. Rev. C}\ }\textbf {\bibinfo {volume} {89}},\ \bibinfo {pages} {064009} (\bibinfo {year} {2014})}\BibitemShut {NoStop}%
\bibitem [{\citenamefont {Wellenhofer}\ \emph {et~al.}(2015)\citenamefont {Wellenhofer}, \citenamefont {Holt},\ and\ \citenamefont {Kaiser}}]{PhysRevC.92.015801}%
  \BibitemOpen
  \bibfield  {author} {\bibinfo {author} {\bibfnamefont {C.}~\bibnamefont {Wellenhofer}}, \bibinfo {author} {\bibfnamefont {J.~W.}\ \bibnamefont {Holt}},\ and\ \bibinfo {author} {\bibfnamefont {N.}~\bibnamefont {Kaiser}},\ }\bibfield  {title} {\bibinfo {title} {Thermodynamics of isospin-asymmetric nuclear matter from chiral effective field theory},\ }\href {https://doi.org/10.1103/PhysRevC.92.015801} {\bibfield  {journal} {\bibinfo  {journal} {Phys. Rev. C}\ }\textbf {\bibinfo {volume} {92}},\ \bibinfo {pages} {015801} (\bibinfo {year} {2015})}\BibitemShut {NoStop}%
\bibitem [{\citenamefont {Lynn}\ \emph {et~al.}(2016)\citenamefont {Lynn}, \citenamefont {Tews}, \citenamefont {Carlson}, \citenamefont {Gandolfi}, \citenamefont {Gezerlis}, \citenamefont {Schmidt},\ and\ \citenamefont {Schwenk}}]{Lynn16QMC3N}%
  \BibitemOpen
  \bibfield  {author} {\bibinfo {author} {\bibfnamefont {J.~E.}\ \bibnamefont {Lynn}}, \bibinfo {author} {\bibfnamefont {I.}~\bibnamefont {Tews}}, \bibinfo {author} {\bibfnamefont {J.}~\bibnamefont {Carlson}}, \bibinfo {author} {\bibfnamefont {S.}~\bibnamefont {Gandolfi}}, \bibinfo {author} {\bibfnamefont {A.}~\bibnamefont {Gezerlis}}, \bibinfo {author} {\bibfnamefont {K.~E.}\ \bibnamefont {Schmidt}},\ and\ \bibinfo {author} {\bibfnamefont {A.}~\bibnamefont {Schwenk}},\ }\bibfield  {title} {\bibinfo {title} {{Chiral Three-Nucleon Interactions in Light Nuclei, Neutron-$\alpha$ Scattering, and Neutron Matter}},\ }\href {https://doi.org/10.1103/PhysRevLett.116.062501} {\bibfield  {journal} {\bibinfo  {journal} {Phys. Rev. Lett.}\ }\textbf {\bibinfo {volume} {116}},\ \bibinfo {pages} {062501} (\bibinfo {year} {2016})}\BibitemShut {NoStop}%
\bibitem [{\citenamefont {Drischler}\ \emph {et~al.}(2016)\citenamefont {Drischler}, \citenamefont {Hebeler},\ and\ \citenamefont {Schwenk}}]{Dris16asym}%
  \BibitemOpen
  \bibfield  {author} {\bibinfo {author} {\bibfnamefont {C.}~\bibnamefont {Drischler}}, \bibinfo {author} {\bibfnamefont {K.}~\bibnamefont {Hebeler}},\ and\ \bibinfo {author} {\bibfnamefont {A.}~\bibnamefont {Schwenk}},\ }\bibfield  {title} {\bibinfo {title} {{Asymmetric nuclear matter based on chiral two- and three-nucleon interactions}},\ }\href {https://doi.org/10.1103/PhysRevC.93.054314} {\bibfield  {journal} {\bibinfo  {journal} {Phys. Rev. C}\ }\textbf {\bibinfo {volume} {93}},\ \bibinfo {pages} {054314} (\bibinfo {year} {2016})}\BibitemShut {NoStop}%
\bibitem [{\citenamefont {Ekstr{\"o}m}\ \emph {et~al.}(2018)\citenamefont {Ekstr{\"o}m}, \citenamefont {Hagen}, \citenamefont {Morris}, \citenamefont {Papenbrock},\ and\ \citenamefont {Schwartz}}]{Ekst17deltasat}%
  \BibitemOpen
  \bibfield  {author} {\bibinfo {author} {\bibfnamefont {A.}~\bibnamefont {Ekstr{\"o}m}}, \bibinfo {author} {\bibfnamefont {G.}~\bibnamefont {Hagen}}, \bibinfo {author} {\bibfnamefont {T.~D.}\ \bibnamefont {Morris}}, \bibinfo {author} {\bibfnamefont {T.}~\bibnamefont {Papenbrock}},\ and\ \bibinfo {author} {\bibfnamefont {P.~D.}\ \bibnamefont {Schwartz}},\ }\bibfield  {title} {\bibinfo {title} {{$\Delta$ isobars and nuclear saturation}},\ }\href {https://doi.org/10.1103/PhysRevC.97.024332} {\bibfield  {journal} {\bibinfo  {journal} {Phys. Rev. C}\ }\textbf {\bibinfo {volume} {97}},\ \bibinfo {pages} {024332} (\bibinfo {year} {2018})}\BibitemShut {NoStop}%
\bibitem [{\citenamefont {Drischler}\ \emph {et~al.}(2019)\citenamefont {Drischler}, \citenamefont {Hebeler},\ and\ \citenamefont {Schwenk}}]{Drischler2019}%
  \BibitemOpen
  \bibfield  {author} {\bibinfo {author} {\bibfnamefont {C.}~\bibnamefont {Drischler}}, \bibinfo {author} {\bibfnamefont {K.}~\bibnamefont {Hebeler}},\ and\ \bibinfo {author} {\bibfnamefont {A.}~\bibnamefont {Schwenk}},\ }\bibfield  {title} {\bibinfo {title} {Chiral interactions up to next-to-next-to-next-to-leading order and nuclear saturation},\ }\href {https://doi.org/10.1103/PhysRevLett.122.042501} {\bibfield  {journal} {\bibinfo  {journal} {Phys. Rev. Lett.}\ }\textbf {\bibinfo {volume} {122}},\ \bibinfo {pages} {042501} (\bibinfo {year} {2019})}\BibitemShut {NoStop}%
\bibitem [{\citenamefont {Carbone}\ and\ \citenamefont {Schwenk}(2019)}]{CarboneSchwenk2019}%
  \BibitemOpen
  \bibfield  {author} {\bibinfo {author} {\bibfnamefont {A.}~\bibnamefont {Carbone}}\ and\ \bibinfo {author} {\bibfnamefont {A.}~\bibnamefont {Schwenk}},\ }\bibfield  {title} {\bibinfo {title} {\textit{Ab initio} constraints on thermal effects of the nuclear equation of state},\ }\href {https://doi.org/10.1103/PhysRevC.100.025805} {\bibfield  {journal} {\bibinfo  {journal} {Phys. Rev. C}\ }\textbf {\bibinfo {volume} {100}},\ \bibinfo {pages} {025805} (\bibinfo {year} {2019})}\BibitemShut {NoStop}%
\bibitem [{\citenamefont {Lu}\ \emph {et~al.}(2020)\citenamefont {Lu}, \citenamefont {Li}, \citenamefont {Elhatisari}, \citenamefont {Lee}, \citenamefont {Drut}, \citenamefont {L\"ahde}, \citenamefont {Epelbaum},\ and\ \citenamefont {Mei\ss{}ner}}]{Lu2020}%
  \BibitemOpen
  \bibfield  {author} {\bibinfo {author} {\bibfnamefont {B.-N.}\ \bibnamefont {Lu}}, \bibinfo {author} {\bibfnamefont {N.}~\bibnamefont {Li}}, \bibinfo {author} {\bibfnamefont {S.}~\bibnamefont {Elhatisari}}, \bibinfo {author} {\bibfnamefont {D.}~\bibnamefont {Lee}}, \bibinfo {author} {\bibfnamefont {J.~E.}\ \bibnamefont {Drut}}, \bibinfo {author} {\bibfnamefont {T.~A.}\ \bibnamefont {L\"ahde}}, \bibinfo {author} {\bibfnamefont {E.}~\bibnamefont {Epelbaum}},\ and\ \bibinfo {author} {\bibfnamefont {U.-G.}\ \bibnamefont {Mei\ss{}ner}},\ }\bibfield  {title} {\bibinfo {title} {{\textit{Ab Initio} Nuclear Thermodynamics}},\ }\href {https://doi.org/10.1103/PhysRevLett.125.192502} {\bibfield  {journal} {\bibinfo  {journal} {Phys. Rev. Lett.}\ }\textbf {\bibinfo {volume} {125}},\ \bibinfo {pages} {192502} (\bibinfo {year} {2020})}\BibitemShut {NoStop}%
\bibitem [{\citenamefont {Keller}\ \emph {et~al.}(2021)\citenamefont {Keller}, \citenamefont {Wellenhofer}, \citenamefont {Hebeler},\ and\ \citenamefont {Schwenk}}]{Keller2021}%
  \BibitemOpen
  \bibfield  {author} {\bibinfo {author} {\bibfnamefont {J.}~\bibnamefont {Keller}}, \bibinfo {author} {\bibfnamefont {C.}~\bibnamefont {Wellenhofer}}, \bibinfo {author} {\bibfnamefont {K.}~\bibnamefont {Hebeler}},\ and\ \bibinfo {author} {\bibfnamefont {A.}~\bibnamefont {Schwenk}},\ }\bibfield  {title} {\bibinfo {title} {{Neutron matter at finite temperature based on chiral effective field theory interactions}},\ }\href {https://doi.org/10.1103/PhysRevC.103.055806} {\bibfield  {journal} {\bibinfo  {journal} {Phys. Rev. C}\ }\textbf {\bibinfo {volume} {103}},\ \bibinfo {pages} {055806} (\bibinfo {year} {2021})}\BibitemShut {NoStop}%
\bibitem [{\citenamefont {Keller}\ \emph {et~al.}(2023)\citenamefont {Keller}, \citenamefont {Hebeler},\ and\ \citenamefont {Schwenk}}]{Keller:2022crb}%
  \BibitemOpen
  \bibfield  {author} {\bibinfo {author} {\bibfnamefont {J.}~\bibnamefont {Keller}}, \bibinfo {author} {\bibfnamefont {K.}~\bibnamefont {Hebeler}},\ and\ \bibinfo {author} {\bibfnamefont {A.}~\bibnamefont {Schwenk}},\ }\bibfield  {title} {\bibinfo {title} {{Nuclear Equation of State for Arbitrary Proton Fraction and Temperature Based on Chiral Effective Field Theory and a Gaussian Process Emulator}},\ }\href {https://doi.org/10.1103/PhysRevLett.130.072701} {\bibfield  {journal} {\bibinfo  {journal} {Phys. Rev. Lett.}\ }\textbf {\bibinfo {volume} {130}},\ \bibinfo {pages} {072701} (\bibinfo {year} {2023})}\BibitemShut {NoStop}%
\bibitem [{\citenamefont {Marino}\ \emph {et~al.}(2024)\citenamefont {Marino}, \citenamefont {Jiang},\ and\ \citenamefont {Novario}}]{Marino:2024tfp}%
  \BibitemOpen
  \bibfield  {author} {\bibinfo {author} {\bibfnamefont {F.}~\bibnamefont {Marino}}, \bibinfo {author} {\bibfnamefont {W.~G.}\ \bibnamefont {Jiang}},\ and\ \bibinfo {author} {\bibfnamefont {S.~J.}\ \bibnamefont {Novario}},\ }\bibfield  {title} {\bibinfo {title} {{Diagrammatic \textit{ab initio} methods for infinite nuclear matter with modern chiral interactions}},\ }\href {https://doi.org/10.1103/PhysRevC.110.054322} {\bibfield  {journal} {\bibinfo  {journal} {Phys. Rev. C}\ }\textbf {\bibinfo {volume} {110}},\ \bibinfo {pages} {054322} (\bibinfo {year} {2024})}\BibitemShut {NoStop}%
\bibitem [{\citenamefont {Tews}\ \emph {et~al.}(2025)\citenamefont {Tews}, \citenamefont {Somasundaram}, \citenamefont {Lonardoni}, \citenamefont {G{\"o}ttling}, \citenamefont {Seutin}, \citenamefont {Carlson}, \citenamefont {Gandolfi}, \citenamefont {Hebeler},\ and\ \citenamefont {Schwenk}}]{Tews:2024owl}%
  \BibitemOpen
  \bibfield  {author} {\bibinfo {author} {\bibfnamefont {I.}~\bibnamefont {Tews}}, \bibinfo {author} {\bibfnamefont {R.}~\bibnamefont {Somasundaram}}, \bibinfo {author} {\bibfnamefont {D.}~\bibnamefont {Lonardoni}}, \bibinfo {author} {\bibfnamefont {H.}~\bibnamefont {G{\"o}ttling}}, \bibinfo {author} {\bibfnamefont {R.}~\bibnamefont {Seutin}}, \bibinfo {author} {\bibfnamefont {J.}~\bibnamefont {Carlson}}, \bibinfo {author} {\bibfnamefont {S.}~\bibnamefont {Gandolfi}}, \bibinfo {author} {\bibfnamefont {K.}~\bibnamefont {Hebeler}},\ and\ \bibinfo {author} {\bibfnamefont {A.}~\bibnamefont {Schwenk}},\ }\bibfield  {title} {\bibinfo {title} {{Neutron matter from local chiral effective field theory interactions at large cutoffs}},\ }\href {https://doi.org/10.1103/r314-6r62} {\bibfield  {journal} {\bibinfo  {journal} {Phys. Rev. Res.}\ }\textbf {\bibinfo {volume} {7}},\ \bibinfo {pages} {033024} (\bibinfo {year} {2025})}\BibitemShut {NoStop}%
\bibitem [{\citenamefont {Alp}\ \emph {et~al.}(2025)\citenamefont {Alp}, \citenamefont {Dietz}, \citenamefont {Hebeler},\ and\ \citenamefont {Schwenk}}]{Alp:2025wjn}%
  \BibitemOpen
  \bibfield  {author} {\bibinfo {author} {\bibfnamefont {F.}~\bibnamefont {Alp}}, \bibinfo {author} {\bibfnamefont {Y.}~\bibnamefont {Dietz}}, \bibinfo {author} {\bibfnamefont {K.}~\bibnamefont {Hebeler}},\ and\ \bibinfo {author} {\bibfnamefont {A.}~\bibnamefont {Schwenk}},\ }\bibfield  {title} {\bibinfo {title} {{Equation~of state and Fermi liquid properties of dense matter based on chiral effective field theory interactions}},\ }\href {https://doi.org/10.1103/ls3l-dn1y} {\bibfield  {journal} {\bibinfo  {journal} {Phys. Rev. C}\ }\textbf {\bibinfo {volume} {112}},\ \bibinfo {pages} {055802} (\bibinfo {year} {2025})}\BibitemShut {NoStop}%
\bibitem [{\citenamefont {Rasmussen}\ and\ \citenamefont {Williams}(2005)}]{Rasmussen2005}%
  \BibitemOpen
  \bibfield  {author} {\bibinfo {author} {\bibfnamefont {C.~E.}\ \bibnamefont {Rasmussen}}\ and\ \bibinfo {author} {\bibfnamefont {C.~K.~I.}\ \bibnamefont {Williams}},\ }\href {https://doi.org/10.7551/mitpress/3206.001.0001} {\emph {\bibinfo {title} {Gaussian Processes for Machine Learning}}}\ (\bibinfo  {publisher} {MIT Press},\ \bibinfo {address} {Cambridge, UK},\ \bibinfo {year} {2005})\BibitemShut {NoStop}%
\bibitem [{\citenamefont {Melendez}\ \emph {et~al.}(2019)\citenamefont {Melendez}, \citenamefont {Furnstahl}, \citenamefont {Phillips}, \citenamefont {Pratola},\ and\ \citenamefont {Wesolowski}}]{Melendez:2019izc_EFT}%
  \BibitemOpen
  \bibfield  {author} {\bibinfo {author} {\bibfnamefont {J.~A.}\ \bibnamefont {Melendez}}, \bibinfo {author} {\bibfnamefont {R.~J.}\ \bibnamefont {Furnstahl}}, \bibinfo {author} {\bibfnamefont {D.~R.}\ \bibnamefont {Phillips}}, \bibinfo {author} {\bibfnamefont {M.~T.}\ \bibnamefont {Pratola}},\ and\ \bibinfo {author} {\bibfnamefont {S.}~\bibnamefont {Wesolowski}},\ }\bibfield  {title} {\bibinfo {title} {{Quantifying Correlated Truncation Errors in Effective Field Theory}},\ }\href {https://doi.org/10.1103/PhysRevC.100.044001} {\bibfield  {journal} {\bibinfo  {journal} {Phys. Rev. C}\ }\textbf {\bibinfo {volume} {100}},\ \bibinfo {pages} {044001} (\bibinfo {year} {2019})}\BibitemShut {NoStop}%
\bibitem [{\citenamefont {Drischler}\ \emph {et~al.}(2020{\natexlab{a}})\citenamefont {Drischler}, \citenamefont {Furnstahl}, \citenamefont {Melendez},\ and\ \citenamefont {Phillips}}]{Drischler:2020hwi_EOS}%
  \BibitemOpen
  \bibfield  {author} {\bibinfo {author} {\bibfnamefont {C.}~\bibnamefont {Drischler}}, \bibinfo {author} {\bibfnamefont {R.~J.}\ \bibnamefont {Furnstahl}}, \bibinfo {author} {\bibfnamefont {J.~A.}\ \bibnamefont {Melendez}},\ and\ \bibinfo {author} {\bibfnamefont {D.~R.}\ \bibnamefont {Phillips}},\ }\bibfield  {title} {\bibinfo {title} {{How Well Do We Know the Neutron-Matter Equation of State at the Densities Inside Neutron Stars? A Bayesian Approach with Correlated Uncertainties}},\ }\href {https://doi.org/10.1103/PhysRevLett.125.202702} {\bibfield  {journal} {\bibinfo  {journal} {Phys. Rev. Lett.}\ }\textbf {\bibinfo {volume} {125}},\ \bibinfo {pages} {202702} (\bibinfo {year} {2020}{\natexlab{a}})}\BibitemShut {NoStop}%
\bibitem [{\citenamefont {Drischler}\ \emph {et~al.}(2020{\natexlab{b}})\citenamefont {Drischler}, \citenamefont {Melendez}, \citenamefont {Furnstahl},\ and\ \citenamefont {Phillips}}]{Drischler:2020yad_matter}%
  \BibitemOpen
  \bibfield  {author} {\bibinfo {author} {\bibfnamefont {C.}~\bibnamefont {Drischler}}, \bibinfo {author} {\bibfnamefont {J.~A.}\ \bibnamefont {Melendez}}, \bibinfo {author} {\bibfnamefont {R.~J.}\ \bibnamefont {Furnstahl}},\ and\ \bibinfo {author} {\bibfnamefont {D.~R.}\ \bibnamefont {Phillips}},\ }\bibfield  {title} {\bibinfo {title} {{Quantifying uncertainties and correlations in the nuclear-matter equation of state}},\ }\href {https://doi.org/10.1103/PhysRevC.102.054315} {\bibfield  {journal} {\bibinfo  {journal} {Phys. Rev. C}\ }\textbf {\bibinfo {volume} {102}},\ \bibinfo {pages} {054315} (\bibinfo {year} {2020}{\natexlab{b}})}\BibitemShut {NoStop}%
\bibitem [{\citenamefont {Entem}\ \emph {et~al.}(2017)\citenamefont {Entem}, \citenamefont {Machleidt},\ and\ \citenamefont {Nosyk}}]{Entem:2017gor}%
  \BibitemOpen
  \bibfield  {author} {\bibinfo {author} {\bibfnamefont {D.~R.}\ \bibnamefont {Entem}}, \bibinfo {author} {\bibfnamefont {R.}~\bibnamefont {Machleidt}},\ and\ \bibinfo {author} {\bibfnamefont {Y.}~\bibnamefont {Nosyk}},\ }\bibfield  {title} {\bibinfo {title} {{High-quality two-nucleon potentials up to fifth order of the chiral expansion}},\ }\href {https://doi.org/10.1103/PhysRevC.96.024004} {\bibfield  {journal} {\bibinfo  {journal} {Phys. Rev. C}\ }\textbf {\bibinfo {volume} {96}},\ \bibinfo {pages} {024004} (\bibinfo {year} {2017})}\BibitemShut {NoStop}%
\bibitem [{\citenamefont {Furnstahl}\ \emph {et~al.}(2015)\citenamefont {Furnstahl}, \citenamefont {Klco}, \citenamefont {Phillips},\ and\ \citenamefont {Wesolowski}}]{Furnstahl:2015rha}%
  \BibitemOpen
  \bibfield  {author} {\bibinfo {author} {\bibfnamefont {R.~J.}\ \bibnamefont {Furnstahl}}, \bibinfo {author} {\bibfnamefont {N.}~\bibnamefont {Klco}}, \bibinfo {author} {\bibfnamefont {D.~R.}\ \bibnamefont {Phillips}},\ and\ \bibinfo {author} {\bibfnamefont {S.}~\bibnamefont {Wesolowski}},\ }\bibfield  {title} {\bibinfo {title} {{Quantifying truncation errors in effective field theory}},\ }\href {https://doi.org/10.1103/PhysRevC.92.024005} {\bibfield  {journal} {\bibinfo  {journal} {Phys. Rev. C}\ }\textbf {\bibinfo {volume} {92}},\ \bibinfo {pages} {024005} (\bibinfo {year} {2015})}\BibitemShut {NoStop}%
\bibitem [{\citenamefont {Hebeler}(2021)}]{Report_Kai}%
  \BibitemOpen
  \bibfield  {author} {\bibinfo {author} {\bibfnamefont {K.}~\bibnamefont {Hebeler}},\ }\bibfield  {title} {\bibinfo {title} {{Three-nucleon forces: Implementation and applications to atomic nuclei and dense matter}},\ }\href {https://doi.org/10.1016/j.physrep.2020.08.009} {\bibfield  {journal} {\bibinfo  {journal} {Phys. Rep.}\ }\textbf {\bibinfo {volume} {890}},\ \bibinfo {pages} {1} (\bibinfo {year} {2021})}\BibitemShut {NoStop}%
\bibitem [{\citenamefont {Bastos}\ and\ \citenamefont {O’Hagan}(2009)}]{Bastos}%
  \BibitemOpen
  \bibfield  {author} {\bibinfo {author} {\bibfnamefont {L.~S.}\ \bibnamefont {Bastos}}\ and\ \bibinfo {author} {\bibfnamefont {A.}~\bibnamefont {O’Hagan}},\ }\bibfield  {title} {\bibinfo {title} {{Diagnostics for Gaussian Process Emulators}},\ }\href {https://doi.org/10.1198/TECH.2009.08019} {\bibfield  {journal} {\bibinfo  {journal} {Technometrics}\ }\textbf {\bibinfo {volume} {51}},\ \bibinfo {pages} {425} (\bibinfo {year} {2009})}\BibitemShut {NoStop}%
\bibitem [{\citenamefont {Baym}\ \emph {et~al.}(1970)\citenamefont {Baym}, \citenamefont {Pethick},\ and\ \citenamefont {Sutherland}}]{BPS}%
  \BibitemOpen
  \bibfield  {author} {\bibinfo {author} {\bibfnamefont {G.}~\bibnamefont {Baym}}, \bibinfo {author} {\bibfnamefont {C.~J.}\ \bibnamefont {Pethick}},\ and\ \bibinfo {author} {\bibfnamefont {P.}~\bibnamefont {Sutherland}},\ }\bibfield  {title} {\bibinfo {title} {{The Ground state of matter at high densities: Equation of state and stellar models}},\ }\href {https://doi.org/10.1086/151216} {\bibfield  {journal} {\bibinfo  {journal} {Astrophys. J.}\ }\textbf {\bibinfo {volume} {170}},\ \bibinfo {pages} {299} (\bibinfo {year} {1970})}\BibitemShut {NoStop}%
\bibitem [{\citenamefont {Keller}\ \emph {et~al.}(2024)\citenamefont {Keller}, \citenamefont {Hebeler}, \citenamefont {Pethick},\ and\ \citenamefont {Schwenk}}]{Keller_2024}%
  \BibitemOpen
  \bibfield  {author} {\bibinfo {author} {\bibfnamefont {J.}~\bibnamefont {Keller}}, \bibinfo {author} {\bibfnamefont {K.}~\bibnamefont {Hebeler}}, \bibinfo {author} {\bibfnamefont {C.~J.}\ \bibnamefont {Pethick}},\ and\ \bibinfo {author} {\bibfnamefont {A.}~\bibnamefont {Schwenk}},\ }\bibfield  {title} {\bibinfo {title} {{Neutron star matter as a dilute solution of protons in neutrons}},\ }\href {https://doi.org/10.1103/PhysRevLett.132.232701} {\bibfield  {journal} {\bibinfo  {journal} {Phys. Rev. Lett.}\ }\textbf {\bibinfo {volume} {132}},\ \bibinfo {pages} {232701} (\bibinfo {year} {2024})}\BibitemShut {NoStop}%
\bibitem [{\citenamefont {Chamel}\ and\ \citenamefont {Haensel}(2008)}]{Chamel_2008}%
  \BibitemOpen
  \bibfield  {author} {\bibinfo {author} {\bibfnamefont {N.}~\bibnamefont {Chamel}}\ and\ \bibinfo {author} {\bibfnamefont {P.}~\bibnamefont {Haensel}},\ }\bibfield  {title} {\bibinfo {title} {{Physics of Neutron Star Crusts}},\ }\href {https://doi.org/10.12942/lrr-2008-10} {\bibfield  {journal} {\bibinfo  {journal} {Living Rev. Rel.}\ }\textbf {\bibinfo {volume} {11}},\ \bibinfo {pages} {10} (\bibinfo {year} {2008})}\BibitemShut {NoStop}%
\bibitem [{\citenamefont {Tews}(2017)}]{Tews_2017}%
  \BibitemOpen
  \bibfield  {author} {\bibinfo {author} {\bibfnamefont {I.}~\bibnamefont {Tews}},\ }\bibfield  {title} {\bibinfo {title} {{Spectrum of shear modes in the neutron-star crust: Estimating the nuclear-physics uncertainties}},\ }\href {https://doi.org/10.1103/PhysRevC.95.015803} {\bibfield  {journal} {\bibinfo  {journal} {Phys. Rev. C}\ }\textbf {\bibinfo {volume} {95}},\ \bibinfo {pages} {015803} (\bibinfo {year} {2017})}\BibitemShut {NoStop}%
\bibitem [{\citenamefont {Steiner}(2008)}]{Steiner_2008}%
  \BibitemOpen
  \bibfield  {author} {\bibinfo {author} {\bibfnamefont {A.~W.}\ \bibnamefont {Steiner}},\ }\bibfield  {title} {\bibinfo {title} {Neutron star inner crust: Nuclear physics input},\ }\href {http://dx.doi.org/10.1103/PhysRevC.77.035805} {\bibfield  {journal} {\bibinfo  {journal} {Phys. Rev. C}\ }\textbf {\bibinfo {volume} {77}},\ \bibinfo {pages} {035805} (\bibinfo {year} {2008})}\BibitemShut {NoStop}%
\bibitem [{\citenamefont {Lattimer}\ \emph {et~al.}(1985)\citenamefont {Lattimer}, \citenamefont {Pethick}, \citenamefont {Ravenhall},\ and\ \citenamefont {Lamb}}]{LLPR}%
  \BibitemOpen
  \bibfield  {author} {\bibinfo {author} {\bibfnamefont {J.~M.}\ \bibnamefont {Lattimer}}, \bibinfo {author} {\bibfnamefont {C.~J.}\ \bibnamefont {Pethick}}, \bibinfo {author} {\bibfnamefont {D.~G.}\ \bibnamefont {Ravenhall}},\ and\ \bibinfo {author} {\bibfnamefont {D.~Q.}\ \bibnamefont {Lamb}},\ }\bibfield  {title} {\bibinfo {title} {Physical properties of hot, dense matter: The general case},\ }\href {https://doi.org/https://doi.org/10.1016/0375-9474(85)90006-5} {\bibfield  {journal} {\bibinfo  {journal} {Nucl. Phys. A}\ }\textbf {\bibinfo {volume} {432}},\ \bibinfo {pages} {646} (\bibinfo {year} {1985})}\BibitemShut {NoStop}%
\bibitem [{\citenamefont {Wang}\ \emph {et~al.}(2021)\citenamefont {Wang}, \citenamefont {Huang}, \citenamefont {Kondev}, \citenamefont {Audi},\ and\ \citenamefont {Naimi}}]{AME}%
  \BibitemOpen
  \bibfield  {author} {\bibinfo {author} {\bibfnamefont {M.}~\bibnamefont {Wang}}, \bibinfo {author} {\bibfnamefont {W.~J.}\ \bibnamefont {Huang}}, \bibinfo {author} {\bibfnamefont {F.~G.}\ \bibnamefont {Kondev}}, \bibinfo {author} {\bibfnamefont {G.}~\bibnamefont {Audi}},\ and\ \bibinfo {author} {\bibfnamefont {S.}~\bibnamefont {Naimi}},\ }\bibfield  {title} {\bibinfo {title} {{The AME 2020 atomic mass evaluation (II). Tables, graphs and references}},\ }\href {https://doi.org/10.1088/1674-1137/abddaf} {\bibfield  {journal} {\bibinfo  {journal} {Chin. Phys. C}\ }\textbf {\bibinfo {volume} {45}},\ \bibinfo {pages} {030003} (\bibinfo {year} {2021})}\BibitemShut {NoStop}%
\bibitem [{\citenamefont {Pethick}\ and\ \citenamefont {Ravenhall}(1995)}]{Pethick1995}%
  \BibitemOpen
  \bibfield  {author} {\bibinfo {author} {\bibfnamefont {C.~J.}\ \bibnamefont {Pethick}}\ and\ \bibinfo {author} {\bibfnamefont {D.~G.}\ \bibnamefont {Ravenhall}},\ }\bibfield  {title} {\bibinfo {title} {{Matter at large neutron excess and the physics of neutron-star crusts}},\ }\href {https://doi.org/10.1146/annurev.ns.45.120195.002241} {\bibfield  {journal} {\bibinfo  {journal} {Annu. Rev. Nucl. Part. Sci.}\ }\textbf {\bibinfo {volume} {45}},\ \bibinfo {pages} {429} (\bibinfo {year} {1995})}\BibitemShut {NoStop}%
\bibitem [{\citenamefont {Haensel}\ \emph {et~al.}(2007)\citenamefont {Haensel}, \citenamefont {Potekhin},\ and\ \citenamefont {Yakovlev}}]{Haensel:2007yy_NS1}%
  \BibitemOpen
  \bibfield  {author} {\bibinfo {author} {\bibfnamefont {P.}~\bibnamefont {Haensel}}, \bibinfo {author} {\bibfnamefont {A.~Y.}\ \bibnamefont {Potekhin}},\ and\ \bibinfo {author} {\bibfnamefont {D.~G.}\ \bibnamefont {Yakovlev}},\ }\href {https://doi.org/10.1007/978-0-387-47301-7} {\emph {\bibinfo {title} {{Neutron Stars~1: Equation of State and Structure}}}}\ (\bibinfo  {publisher} {Springer},\ \bibinfo {address} {New York, NY},\ \bibinfo {year} {2007})\BibitemShut {NoStop}%
\bibitem [{\citenamefont {Grams}\ \emph {et~al.}(2022{\natexlab{a}})\citenamefont {Grams}, \citenamefont {Somasundaram}, \citenamefont {Margueron},\ and\ \citenamefont {Reddy}}]{Grams_2022_2}%
  \BibitemOpen
  \bibfield  {author} {\bibinfo {author} {\bibfnamefont {G.}~\bibnamefont {Grams}}, \bibinfo {author} {\bibfnamefont {R.}~\bibnamefont {Somasundaram}}, \bibinfo {author} {\bibfnamefont {J.}~\bibnamefont {Margueron}},\ and\ \bibinfo {author} {\bibfnamefont {S.}~\bibnamefont {Reddy}},\ }\bibfield  {title} {\bibinfo {title} {{Properties of the neutron star crust: Quantifying and correlating uncertainties with improved nuclear physics}},\ }\href {https://doi.org/10.1103/PhysRevC.105.035806} {\bibfield  {journal} {\bibinfo  {journal} {Phys. Rev. C}\ }\textbf {\bibinfo {volume} {105}},\ \bibinfo {pages} {035806} (\bibinfo {year} {2022}{\natexlab{a}})}\BibitemShut {NoStop}%
\bibitem [{\citenamefont {Grams}\ \emph {et~al.}(2022{\natexlab{b}})\citenamefont {Grams}, \citenamefont {Margueron}, \citenamefont {Somasundaram}, \citenamefont {Chamel},\ and\ \citenamefont {Goriely}}]{Grams:2022ojr}%
  \BibitemOpen
  \bibfield  {author} {\bibinfo {author} {\bibfnamefont {G.}~\bibnamefont {Grams}}, \bibinfo {author} {\bibfnamefont {J.}~\bibnamefont {Margueron}}, \bibinfo {author} {\bibfnamefont {R.}~\bibnamefont {Somasundaram}}, \bibinfo {author} {\bibfnamefont {N.}~\bibnamefont {Chamel}},\ and\ \bibinfo {author} {\bibfnamefont {S.}~\bibnamefont {Goriely}},\ }\bibfield  {title} {\bibinfo {title} {{Neutron star crust properties: comparison between the compressible liquid-drop model and the extended Thomas-Fermi approach}},\ }\href {https://doi.org/10.1088/1742-6596/2340/1/012030} {\bibfield  {journal} {\bibinfo  {journal} {J. Phys. Conf. Ser.}\ }\textbf {\bibinfo {volume} {2340}},\ \bibinfo {pages} {012030} (\bibinfo {year} {2022}{\natexlab{b}})}\BibitemShut {NoStop}%
\bibitem [{\citenamefont {Svensson}\ \emph {et~al.}(2026)\citenamefont {Svensson}, \citenamefont {Tichai}, \citenamefont {Hebeler},\ and\ \citenamefont {Schwenk}}]{Svensson:2025jde}%
  \BibitemOpen
  \bibfield  {author} {\bibinfo {author} {\bibfnamefont {I.}~\bibnamefont {Svensson}}, \bibinfo {author} {\bibfnamefont {A.}~\bibnamefont {Tichai}}, \bibinfo {author} {\bibfnamefont {K.}~\bibnamefont {Hebeler}},\ and\ \bibinfo {author} {\bibfnamefont {A.}~\bibnamefont {Schwenk}},\ }\bibfield  {title} {\bibinfo {title} {{Bayesian approach for many-body uncertainties in nuclear structure: Many-body perturbation theory for finite nuclei}},\ }\href {https://doi.org/10.1103/y8kt-mgf5} {\bibfield  {journal} {\bibinfo  {journal} {Phys. Rev. C}\ }\textbf {\bibinfo {volume} {113}},\ \bibinfo {pages} {024303} (\bibinfo {year} {2026})}\BibitemShut {NoStop}%
\bibitem [{\citenamefont {Rutherford}\ \emph {et~al.}(2024)\citenamefont {Rutherford}, \citenamefont {Mendes}, \citenamefont {Svensson}, \citenamefont {Schwenk}, \citenamefont {Watts}, \citenamefont {Hebeler}, \citenamefont {Keller}, \citenamefont {Prescod-Weinstein}, \citenamefont {Choudhury}, \citenamefont {Raaijmakers} \emph {et~al.}}]{Rutherford:2024srk}%
  \BibitemOpen
  \bibfield  {author} {\bibinfo {author} {\bibfnamefont {N.}~\bibnamefont {Rutherford}}, \bibinfo {author} {\bibfnamefont {M.}~\bibnamefont {Mendes}}, \bibinfo {author} {\bibfnamefont {I.}~\bibnamefont {Svensson}}, \bibinfo {author} {\bibfnamefont {A.}~\bibnamefont {Schwenk}}, \bibinfo {author} {\bibfnamefont {A.~L.}\ \bibnamefont {Watts}}, \bibinfo {author} {\bibfnamefont {K.}~\bibnamefont {Hebeler}}, \bibinfo {author} {\bibfnamefont {J.}~\bibnamefont {Keller}}, \bibinfo {author} {\bibfnamefont {C.}~\bibnamefont {Prescod-Weinstein}}, \bibinfo {author} {\bibfnamefont {D.}~\bibnamefont {Choudhury}}, \bibinfo {author} {\bibfnamefont {G.}~\bibnamefont {Raaijmakers}}, \emph {et~al.},\ }\bibfield  {title} {\bibinfo {title} {{Constraining the Dense Matter Equation of State with New NICER Mass{\textendash}Radius Measurements and New Chiral Effective Field Theory Inputs}},\ }\href {https://doi.org/10.3847/2041-8213/ad5f02} {\bibfield  {journal} {\bibinfo  {journal} {Astrophys. J. Lett.}\ }\textbf {\bibinfo {volume}
  {971}},\ \bibinfo {pages} {L19} (\bibinfo {year} {2024})}\BibitemShut {NoStop}%
\bibitem [{\citenamefont {Göttling}\ \emph {et~al.}(2026)\citenamefont {Göttling}, \citenamefont {Hoff}, \citenamefont {Hebeler},\ and\ \citenamefont {Schwenk}}]{zenodo}%
  \BibitemOpen
  \bibfield  {author} {\bibinfo {author} {\bibfnamefont {H.}~\bibnamefont {Göttling}}, \bibinfo {author} {\bibfnamefont {L.}~\bibnamefont {Hoff}}, \bibinfo {author} {\bibfnamefont {K.}~\bibnamefont {Hebeler}},\ and\ \bibinfo {author} {\bibfnamefont {A.}~\bibnamefont {Schwenk}},\ }\bibfield  {title} {\bibinfo {title} {Data: Neutron star crust and outer core equation of state from chiral effective field theory with quantified uncertainties}\ }\href {https://doi.org/10.5281/zenodo.18983106} {10.5281/zenodo.18983106} (\bibinfo {year} {2026})\BibitemShut {NoStop}%
\end{thebibliography}%

\end{document}